\documentclass[prd,aps,floatfix,superscriptaddress,nofootinbib,twocolumn,10pt,English]{revtex4}

\usepackage{amssymb,amsmath}
\usepackage{graphicx}
\usepackage[percent]{overpic}
\usepackage{overpic}
\usepackage{color}
\usepackage[colorlinks=true]{hyperref}

\begin{document}  

\title{Evolution of highly eccentric binary neutron stars including tidal effects}

\author{ Huan Yang}
\affiliation{University of Guelph, Guelph, Ontario N2L3G1, Canada}
\affiliation{Perimeter Institute for Theoretical Physics, Waterloo, Ontario N2L2Y5, Canada}
\author{William E. East}
\affiliation{Perimeter Institute for Theoretical Physics, Waterloo, Ontario N2L2Y5, Canada}
\author{Vasileios Paschalidis}
\affiliation{Theoretical Astrophysics Program, Departments of Astronomy and Physics, University of Arizona, Tucson, Arizona 85721}
\author{Frans Pretorius}
\affiliation{Department of Physics, Princeton University, Princeton, New Jersey 08544, USA.}
\affiliation{CIFAR, Cosmology \& Gravity Program, Toronto, Ontario M5G 1Z8, Canada}
\author{Raissa F. P. Mendes}
\affiliation{Instituto de F\'isica, Universidade Federal Fluminense, Niter\'oi, 24210-346, Rio de Janeiro, Brazil}

\begin{abstract} 
This work is the first in a series of studies aimed at understanding the
dynamics of highly eccentric binary neutron stars, and constructing
an appropriate gravitational-waveform model for detection. Such binaries
are possible sources for ground-based gravitational wave detectors, and are
expected to form through dynamical scattering and multi-body interactions
in globular clusters and galactic nuclei. In contrast to black holes,
oscillations of neutron stars are generically excited by tidal effects
after close pericenter passage.  Depending on the equation of state,
this can enhance the loss of orbital energy by up to tens of percent 
over that radiated away by gravitational waves 
during an orbit. Under the same interaction mechanism, part of the orbital
angular momentum is also transferred to the star. We calculate the impact
of the neutron star oscillations on the orbital evolution of such systems,
and compare these results to full numerical simulations.  Utilizing a
Post-Newtonian flux description we propose a preliminary model to predict
the timing of different pericenter passages. A refined version of this
model (taking into account Post-Newtonian corrections to the tidal coupling
and the oscillations of the stars) may serve as a waveform model for such
highly eccentric systems.
\end{abstract}

\maketitle

\section{Introduction}

The LIGO (Laser Interferometric Gravitational Wave Observatory) and
Virgo collaborations have already detected five binary black hole (BH)
coalescence events
\cite{Abbott:2016blz,Abbott:2016nmj,PhysRevLett.118.221101,abbott2017gw170608,abbott2017gw170814}
and one binary neutron star (NS) event
\cite{abbott2017gw170817}. Current event rates indicate that as LIGO
and Virgo reach design sensitivity in the next few years, many more
binary NS coalescences will be observed~\cite{abbott2017gw170817}.
The formation of binary compact object (CO) systems can be broadly classified
into two main categories: field binaries and dynamically assembled (cluster)
binaries. In the first channel, the progenitor is a stellar
binary where both stars eventually form COs
following stellar collapse (with or without
supernovae). Such a binary may gain a sizable amount of eccentricity
through a supernovae kick, however subsequent gravitational wave (GW)
emission will reduce the eccentricity with time, leading to a nearly
circular orbit by the time the emission enters the LIGO band.
In the second channel, CO binaries form through
various N-body interactions that take place in dense stellar
environments, including dynamical capture 
\cite{o2009gravitational,lee2010,kocsis2012repeated,fabian1975tidal,pooley2003dynamical},
and exchange interactions during binary-single and binary-binary
interactions\cite{Samsing:2013kua,Rodriguez:2015oxa,samsing2017eccentric}.

An exciting aspect of dynamically assembled binaries is that some
fraction of these systems will emit GWs in the LIGO/Virgo bands while
their orbits are highly eccentric (for order-of-magnitude estimates of
these fractions see
e.g.~\cite{East:2012xq,Samsing:2013kua,samsing2017eccentric}).
A further mechanism that can produce eccentric mergers within ground-based GW detector bands
are hierarchical triple systems where a Kozai-Lidov type resonance
occurs~\cite{2003ApJ...598..419W,2013PhRvL.111f1106S,naoz2016eccentric,2017ApJ...841...77A,2018arXiv180508212R}; 
such systems could form both in field and cluster environments, though
for our purposes of studying eccentric mergers we will group them
in the ``dynamically assembled'' category.
Currently, the rate estimates for dynamically assembled binaries are
quite uncertain, but it is possible that they contribute a
non-negligible fraction of events detectable by LIGO and future
ground-based GW detectors.

If a highly eccentric binary system contains
one or more neutron stars it will also have distinct phenomenology.
For example, the star will be tidally perturbed after each pericenter
passage \cite{1977ApJ...216..914T}, leaving a weak, oscillatory GW
imprint as a tail of the main GW burst produced at pericenter
\cite{gold2012eccentric,East:2011xa,East:2012ww}, along with possible
electromagnetic emission due to crust
cracking (if the pericenter distance is sufficiently small) \cite{East:2012ww,Tsang:2013mca}.  
Probing these post-burst NS oscillations in GWs offers an opportunity
to study the oscillation modes of cold, perturbed neutron stars (NS
asteroseismology), which are distinct from the modes of a post-merger
hypermassive NS remnant (see
e.g.~\cite{Paschalidis:2016vmz,Baiotti:2016qnr} for recent reviews on
this topic)~\footnote{Some have proposed that quasi periodic
  oscillations of magnetars are associated with their crustal modes
  \cite{barat1983fine,duncan1998global,0004-637X-841-1-54}, which
  could be another way to realize NS asteroseismology.}.  Therefore it
is interesting to study the dynamics of such binaries and develop an
appropriate GW detection template that includes both the burst and
post-burst phenomenology. However, we point out that as these
post-burst oscillations are generally much weaker than the pericenter
bursts, direct observation of the accompanying GWs may require at
least third-generation ground-based detectors and appropriate data
analysis methods, such as coherent mode
stacking~\cite{yang2017black,PhysRevD.97.024049,berti2018extreme} (see
also~\cite{Bose:2017jvkxs,Brito:2018rfr}).

In this work we explicitly compute the amount of energy/angular
momentum deposited into the star(s) due to dynamical tidal
excitations. We find that during a close passage, the change of
orbital energy due to the tidal interaction may be up
to tens of percent of the energy carried away by GWs during the same time,
depending on the equation of state (EOS) of the
star. This observation is consistent with a recent study on f-mode
excitation in BH-NS binaries \cite{parisi2018gravitational}, but
different from the conclusion drawn in
\cite{chirenti2016gravitational}, which, as we will discuss in Sec.~\ref{sec2}, 
we attribute to the incomplete
summation of modes used in \cite{chirenti2016gravitational}.
We also find that a non-negligible amount
of orbital angular momentum is transferred to the star, mainly through
f-mode excitations. These f-modes decay with time due to GW emission,
with a quality factor that is generally very high, so that a
significant fraction of the mode energy can remain by the next
pericenter passage. Therefore an accurate model for the orbital
evolution must include the evolution of the star's f-modes.

To  model the orbit, we adopt the osculating-orbit
approximation used in \cite{loutrel2017eccentric} for describing the
orbital evolution of highly eccentric binary black holes   (see
for example~\cite{Huerta:2016rwp} for a model of lower eccentricity
binaries). Within such an approximation, each orbital cycle is
described by an eccentric orbit with fixed energy, angular momentum,
and eccentricity in the Post-Newtonian (PN) expansion, but these quantities
change from one cycle to the next according to the accumulated flux calculated
within a cycle. By definition, such an approximation fails if the conserved quantities change significantly within an orbital cycle. For the GW flux generated by the tidal excitations, we
only consider Newtonian order terms of the star(s)'s oscillations
under the tidal field of the companion. Including higher PN tidal
effects may require dealing with nontrivial gauge issues
connecting single star calculations to a consistent binary NS
calculation~\cite{gralla2018ambiguity}. For the flux generated by the
orbital motion, we have used the 3PN flux formulas in
\cite{blanchet2014gravitational, loutrel2017eccentric}.

This paper is organized as follows. In Sec.~\ref{sec2} we 
calculate the excitation of NS modes due to tidal interactions, and
their influence on the orbital evolution by changing the orbital energy and
angular momentum. In Sec.~\ref{sec3} we apply these results to polytropic stars,
and compare the predictions with a full numerical calculation in Sec.~\ref{sec4}. We
review the PN description of eccentric orbits, and propose a model to describe
the timing of sequential bursts of eccentric binary NS systems in
Sec.~\ref{sec5}. 
Unless otherwise noted, we use geometric units with $G=c=1$ throughout.

\section{Formalism}\label{sec2}

\subsection{Mode excitation, energy and angular momentum transfer}

Let us consider f-mode excitation of a NS due to the tidal field
generated by its companion star. Other eigenmodes
(such as p-modes) of the star are also excited by the tidal field, but
their contribution to the total modal energy is much smaller than the
f-modes \cite{press1977formation}, so in this work we focus only on
f-modes. The generalization of the calculation performed here to
additional modes is straightforward. In addition, we shall only
perform the analysis of the stellar deformation to the leading,
Newtonian order, because the tidal-induced energy and angular momentum
transfer are already of higher PN order compared to the GW radiation
back-reaction.  We shall also only keep the leading order, quadrupole
piece of the tidal field, as contributions from higher multipoles are
negligible \cite{press1977formation}.

According to \cite{press1977formation}, the time domain equation of motion can be written as
\begin{align}\label{eqdy}
\mathcal{L}\,{\bf \xi} = \nabla U\,,
\end{align}
where 
$\xi$ is the Lagrangian displacement vector field, $U$ is the tidal potential 
and $\mathcal{L}$ is a self-adjoint operator whose detailed form is not important here. 
Let us denote the eigenfrequency of each mode as $\omega_n$ and the eigenfunction as ${\bf \xi}^{(n)}$,
where $n$ labels the spherical harmonic indices $(l,m)$. 
We define an  inner product as
 \begin{align}
 \langle \chi | \eta \rangle = \int d^3 x\, \rho\, \chi^*\, \eta\,,
 \end{align}
 where $\rho$ is the rest-mass density.
 The mode eigenfunctions are normalized such that $\langle {\bf \xi}^{(n)} | {\bf \xi}^{(n')} \rangle =\delta_{n\,n'}$.
 The Lagrangian displacement field can be decomposed as
\begin{align}\label{eqdecom}
 \xi_i =\sum_{n =(l,m)} \left [A_n(t)\,\xi^{(n)}_i+h.c. \right ]\,,\quad i=x,y,z,
 \end{align}
where $h.c.$ stands for hermitian conjugate.
As the pericenter passage timescale is much shorter than the mode decay
timescale (due to GW radiation and viscous damping), and restricting to
non-rotating stars as in \cite{chirenti2016gravitational,press1977formation},
$\mathcal{L}$ 
should act solely as a second derivative when applied to functions of time.
We plug Eq.~\eqref{eqdecom} into
Eq.~\eqref{eqdy} and take the inner product with $\xi^{(n)}$ 
to obtain the time domain mode evolution equation as 
\begin{align}\label{eqmodeeom} 
\ddot A^{(n)}+\omega^2_n A^{(n)} = \langle {\bf \xi}^{(n)} | \nabla U \rangle\,.
\end{align}
 
We focus on the f-modes of NSs, which can be intuitively understood as the fundamental oscillation
modes of the star, and ignore the less relevant p or g overtone modes.
The eigenfunction ${\bf \xi}^{(n)}$ can be expressed in vector spherical harmonics:
\begin{align}\label{eqeigenwf} 
{\bf \xi}^{(n)} = \left ( \xi^{(n)}_R(r) {\bf
e_r}+\xi^{(n)}_S(r) r \nabla \right ) Y_{lm}(\theta,\phi)\,, 
\end{align} 
where the determination of $\xi^{(n)}_R$ and $\xi^{(n)}_S$ is
discussed in Appendix \ref{appendixA}. Note that by writing down the
mode decomposition in the time domain in Eq.~\eqref{eqdecom}, we need
to take into account the double counting problem, as each bracket in
the summation includes both $\pm m$ modes (with $m$ running from $-l$
to $l$). One way to resolve this problem is to only include modes
with $m\ge 0$ in the summation, and in particular remove the Hermitian
conjugate terms for $m=0$ modes. Another way is to allow a complex
displacement field for each mode, and remove all Hermitian conjugate
terms in the summation. The reality condition for ${\bf \xi}$ will be
enforced by the fact that $A^{l,m} =A^{l,-m *}$. These two methods
give the same result, and we shall adopt the second approach because
it is more convenient for both the time domain and the frequency
domain analyses.

The quadrupole piece of the tidal potential can be expressed as
\begin{align}
U = \frac{\mathcal{E}_{ij} x^i x^j}{2}\,,
\end{align}
with $\mathcal{E}_{ij}$ being the electric part of the tidal tensor \cite{poisson2010geometry,yang2017general}.
During a pericenter passage, the energy deposited into the star is given by \cite{press1977formation} (in the Lagrange description, $x_j \rightarrow x_j+\xi_j$, and we neglect $\mathcal{O}(\xi^2)$ terms)
\begin{align}\label{eqemode}
\Delta E_{\rm mode} & =\int dt\, \int d^3 x \rho \, {\bf v} \cdot \nabla U \,
  = \int dt\,\int d^3 x \rho \, v^i \nabla_i U \,\nonumber \\
&  \approx \int dt\,  \int d^3 x \rho \, \mathcal{E}_{ik}(t) \,\dot{\xi}^i\, x^k \nonumber \\
  & = \sum_n \int dt\, \mathcal{E}^{ik}(t)\, \dot{A}^{(n)}(t) \int d^3 x\, \rho\, \xi^{(n)}_i(x) x_k\,.
  \end{align}

On the one hand, we  note that Eq.~\eqref{eqmodeeom} (within the quadrupole approximation) can be written as 
\begin{align}\label{eqenit}
\ddot A^{(n)}+\omega^2_n A^{(n)} &=  \int d^3 x\, \rho\, \xi^{(n)*}_i(x) x_k \,\mathcal{E}^{ik}(t) \nonumber\\
& \equiv \mathcal{O}^*_{ik} \mathcal{E}^{ik}(t)\,
\end{align}
so that the change of the mode energy  becomes
\begin{align}\label{eqem}
 \Delta E_{\rm mode} & = \sum_n \int dt\, \mathcal{E}^{ik}(t)\, \dot{A}^{(n)}(t) \mathcal{O}_{ik}  \nonumber \\
 & =\sum_n \int dt\, \dot{A}^{(n)}(t) (\ddot{A}^{(n)*}+\omega_n^2 A^{(n)*})  \nonumber \\
 &=\Delta \left \{\frac{1}{2} \sum_n [ |\dot{A}^{(n)}(t) |^2+\omega^2_n |A^{(n)}|^2 ]\right \}\,.
    \end{align} 
The physical meaning of the above expression is obvious: the change in mode energy  can be divided into kinetic energy change and potential energy change. 

As an initial value problem, the solution of Eq.~(\ref{eqenit}) is given by
\begin{align}\label{eqat}
A^{(n)}(t) = &\frac{1}{\omega_n}\int^t_{-\infty} d t'\,\sin [\omega_n (t-t')] \mathcal{E}_{ij}(t') \mathcal{O}^{ij*} \nonumber \\
&+\left (A^{(n)}_{\rm init}|_{t=0} \cos{\omega_n t} +\frac{\dot{A}^{(n)}_{\rm init}|_{t=0} }{\omega_n} \sin{\omega_n t}\right )\,,
\end{align}
with $A^{n}_{\rm init}(t)$ being an initial oscillation in the given mode of the star prior to any tidal interaction.
Within the osculating orbit approximation, stringing together a sequence of pericenter encounters, 
the new tidally induced oscillation at each encounter can be modeled by an expression of the form (\ref{eqat}),
with the constants $(A^{(n)}_{\rm init}|_{t=0}, \dot{A}^{(n)}_{\rm init}|_{t=0})$ determined 
using the previous cycle's $A^{(n)}(t)$, appropriately damped by GW emission.

On the other hand, we can rewrite the mode evolution equation in the frequency domain, with
\begin{align}\label{eqfreq}
\tilde{A}^{(n)} =\frac{\langle {\bf \xi}^{(n)} | \nabla \tilde{U} \rangle}{\omega^2_n-\omega^2}=\frac{\mathcal{O}^*_{ik} \tilde{\mathcal{E}}^{ik}}{\omega^2_n-\omega^2}\,,
\end{align}
and
\begin{align}
\chi(t) \equiv \int^\infty_{-\infty} d \omega \, \tilde{\chi}(\omega )\, e^{-i \omega t}\,.
\end{align}
The frequency domain equation (Eq.~\eqref{eqfreq}) is suitable to describe the evolution 
with zero initial oscillations, e.g., in cases where the mode decay timescale is shorter than 
the orbital timescale. 
In the frequency domain description, the energy deposited into modes is \cite{press1977formation}
\begin{align}
\Delta E_{\rm mode} =2 \pi^2 \sum_n |\mathcal{O}^*_{ik} \tilde{\mathcal{E}}^{ik}(\omega_n)|^2\,.
\end{align}

In addition to energy, part of the orbital angular momentum is also
transferred to the star through the dynamical coupling between the
tidal bulge and the companion star. Similar to Eq.~\eqref{eqemode}, with the power injected replaced by the torque acting on each mass element, we have 
\begin{align}\label{eqangc}
\Delta {J_{\rm mode }}^i  &= \int dt \int d^3 x \, \rho\,\epsilon^{ijk} (x_j+\xi_j) \nabla_k U\,\nonumber \\
& = \int dt \epsilon^{ijk} {\mathcal{E}_{k}}^l\,\int d^3 x\,\rho\, (x_j+\xi_j)(x_l+\xi_l) \nonumber \\
&\approx \int dt \epsilon^{ijk} {\mathcal{E}_{k}}^l\,\int d^3 x\,\rho\, (x_j \xi_l+\xi_j x_l) \nonumber \\
& =2\sum_n \int dt\, {\mathcal{E}_{k}}^l(t)\, A^{(n)}(t) \int d^3 x\, \rho\, \xi^{(n)}_{\langle l}(x) x_{j\rangle }\, \epsilon^{ijk} \nonumber \\
&=2 \sum_n \int dt\, \epsilon^{ijk}\, {\mathcal{E}_{k}}^l(t)\, A^{(n)}(t) \mathcal{O}_{\langle jl \rangle} \,,
\end{align}
with the symmetrized tensor $\mathcal{O}_{\langle jl \rangle}$ being $(\mathcal{O}_{jl}+\mathcal{O}_{lj})/2$, and in the third line we have neglected $\mathcal{O}(\xi^2)$ terms. In this study, we choose the coordinate system such that the stellar trajectory resides on the equatorial plane, and only  consider the change of angular momentum along the z-axis:

\begin{align}
\Delta J_{\rm mode,z} = 2\sum_n \int dt\, \,\left( \mathcal{E}^{y l}(t) \mathcal{O}_{\langle x l \rangle}-\mathcal{O}_{\langle y l \rangle} \mathcal{E}^{x l}(t)\right)\, A^{(n)}(t) \,.
\end{align}

In the frequency domain, similarly to the evaluation for the mode energy, Eq.~\eqref{eqangc} can be rewritten as
 \begin{align}
\Delta J_{\rm mode, i} & = 4 \pi \sum_n \int d\omega\, {\tilde{\mathcal{E}}^{kl *}}(\omega)\, \tilde{A}^{(n)}(\omega)\mathcal{O}_{\langle lj \rangle} \epsilon_{ijk} \nonumber \\
 =& 4 \pi \sum_n \int d \omega \, {\tilde{\mathcal{E}}^{kl *}}(\omega)\, \mathcal{O}_{\langle lj \rangle} \epsilon_{ijk}\frac{ \tilde{\mathcal{E}}^{gh } \mathcal{O}^*_{hg} }{\omega^2_n-\omega^2-i\epsilon}\,\nonumber \\
=& \frac{4 \pi^2 i}{\omega_n} \sum_n \epsilon_{ijk} [\tilde{\mathcal{E}}^{kl*}(\omega_n)\tilde{\mathcal{E}}^{gh } (\omega_n) -\tilde{\mathcal{E}}^{kl}(\omega_n)\tilde{\mathcal{E}}^{gh*} (\omega_n)]\nonumber \\
&\times \mathcal{O}^*_{\langle hg \rangle} \mathcal{O}_{\langle lj \rangle}\,,
\end{align}
where we have used the fact that $\mathcal{E}$ is a symmetric tensor.

\subsection{Typical orbits}

The energy and angular momentum lost to mode
excitations is generally smaller than that lost to GW radiation from
the orbital motion. As a result, we adopt the Newtonian description of
eccentric orbits when we compute the NS mode excitations. Such a
simplified treatment also facilitates a direct comparison with previous
work~\cite{chirenti2016gravitational,press1977formation}. On the
other hand, to track the long-term orbital evolution, a
PN/quasi-Keplerian formalism \cite{blanchet2014gravitational} becomes
necessary, which we discuss in Sec.~\ref{sec5}. In this section, we
discuss marginally unbound (parabolic) and bound eccentric orbits separately.

\subsubsection{Parabolic orbit}\label{sec:hyper}

In the rest frame of a reference star,  the parabolic orbit of the companion star can be parameterized by
\begin{align}\label{eqntrajectory}
 r_t=& R_0(1+\tau^2)\,,\nonumber \\
 t= &\left [ \frac{2R^3_0}{G(M_*+M)}\right ]^{1/2}(\tau+\tau^3/3)\,,\nonumber \\
 \tau=&\tan \frac{\Phi}{2}\,,
 \end{align}
 where $r_t$ is the orbital separation, $t$ the time from
   pericenter passage, $\Phi$ the true anomaly, $M$ and $M_*$ are the
 masses of the reference and incoming star, respectively, and $R_0$ is
 the pericenter distance. In Cartesian coordinates, the position in
 $(x,y)$ (equatorial plane) is then given by
 \begin{align}
 x=r_t \cos \Phi,\quad y=r_t \sin \Phi\,.
 \end{align}
 
 The quadrupole tidal tensor, generated by an incoming star following the above trajectory is given by
  \begin{align}
 \mathcal{E}= \frac{M_*}{r_t^3}\left [ \begin{array}{ccc}
 -\frac{1}{2}-\frac{3}{2}\cos 2\Phi & \frac{3}{2}  \sin 2\Phi & 0\\
 \frac{3}{2}  \sin 2\Phi & \frac{-1}{2}+\frac{3}{2} \cos 2\Phi & 0 \\
 0 & 0 & 1
   \end{array}
 \right ]
 \end{align}

The symmetrized overlap tensor $\mathcal{O}_{\langle \rangle}$ is given by
 \begin{align}
 \mathcal{O}_{\langle a b \rangle}=\frac{ Q^{(n)}_\xi}{2} \int d \Omega (e_{\hat a} \cdot e_{\hat r}) (e_{\hat b} \cdot e_{\hat r}) Y_{lm}\,,
 \end{align} 
 with $e_{\hat a}$ being a unit coordinate vector, $e_{\hat r}$ being the unit radial vector and $Q^{(n)}_\xi$ defined as ($R_*$ being the radius of the star) 
 \begin{align}
 Q^{(n)}_\xi \equiv 2\int^{R_*}_0 dr\,r^3\, \rho (\xi^{(n)}_R+3\xi^{(n)}_S)\,.
 \end{align}
 
 The quadrupole tidal field excites NS modes with $l=2$ only. Summing
 over the azimuthal wave numbers of the f-modes (and neglecting
 contributions from p, g-modes), in the time domain, the angular
 momentum shift of the star is 
 \begin{widetext}
  \begin{align}
\Delta J_{\rm mode, z} &= 2\sum_{l=2,m} \int dt\, \,( \mathcal{E}^{y l}(t) \mathcal{O}_{x l}-\mathcal{O}_{y l} \mathcal{E}^{x l}(t))\, A^{(n)}(t) \nonumber \\
& =  \sqrt{6 \pi/5} \int dt\, \frac{M_*}{r_t^3} [- i  Q^{22}_\xi e^{ 2 i \Phi}  A_{22}(t)+i Q^{2,-2}_\xi e^{ -2 i \Phi} A_{2,-2}(t)]\,, 
\end{align}
where
\begin{align}\label{eqamap}
A_{22} & =\frac{Q^{22}_\xi}{2\omega_{22}}\int^t_{-\infty} dt'\, \sqrt{6 \pi/5} \frac{M_*}{r_t^3} \sin \omega_{22} (t-t') e^{-2 i \Phi(t')}\nonumber \\
&+\left (A^{(22)}_{\rm init}|_{t=0} \cos{\omega_{22} t} +\frac{\dot{A}^{(22)}_{\rm init}|_{t=0} }{\omega_{22}} \sin{\omega_{22} t}\right )\,,\nonumber \\
A_{2,-2} & =\frac{Q^{2,-2}_\xi}{2\omega_{2,-2}}\int^t_{-\infty} dt'\, \sqrt{6 \pi/5} \frac{M_*}{r_t^3} \sin \omega_{2,-2} (t-t') e^{2 i \Phi(t')} \nonumber \\
&+\left (A^{(2,-2)}_{\rm init}|_{t=0} \cos{\omega_{2,-2} t} +\frac{\dot{A}^{(2,-2)}_{\rm init}|_{t=0} }{\omega_{2,-2}} \sin{\omega_{2,-2} t}\right )\,.
\end{align}

For non-rotating stars, or if we neglect the split of oscillation
modes due to rotation, we have $\omega_{22}=\omega_{2,-2}$,
$Q^{l,m}_\xi=Q^{l,-m*}_\xi$ and $A_{22}=A^*_{2,-2}$:
\begin{align}\label{eqdjhyper}
\Delta J_{\rm mode, z} &=  |Q^{22}_\xi|^2\frac{3 \pi}{5 \omega_{22}} \int^\infty_{-\infty} dt\,\int^t_{-\infty} dt'\, \frac{M^2_*}{r_t^3 {r_t'}^3} \sin \omega_{22}(t-t') \sin [2(\Phi-\Phi')]\,\nonumber \\
& =  |Q^{22}_\xi|^2 \frac{3 \pi}{10 \omega_{22}} \int^\infty_{-\infty} dt\,\int^\infty_{-\infty} dt'\, \frac{M^2_*}{r_t^3 {r_t'}^3} \sin \omega_{22}(t-t') \sin [2(\Phi-\Phi')]\,.
\end{align} 

The energy change under the same assumption (no initial oscillation) can be obtained by using Eq.~\eqref{eqemode} and \eqref{eqat}:
\begin{align}\label{eqehyper}
\Delta E_{\rm mode} &=\sum_n \int^\infty_{-\infty} dt \int^t_{-\infty} \cos \omega_{n}(t-t') \mathcal{E}_{ij}(t') \mathcal{O}^{ij*} \mathcal{E}_{pq}(t) \mathcal{O}^{pq} \nonumber \\
& =  |Q^{22}_\xi|^2 \frac{3 \pi}{10} \int^\infty_{-\infty} dt\,\int^\infty_{-\infty} dt'\, \frac{M^2_*}{r_t^3 {r_t'}^3} \cos \omega_{22}(t-t') \cos [2(\Phi-\Phi')] \nonumber\\
&+|Q^{20}_\xi|^2 \frac{ \pi}{10} \int^\infty_{-\infty} dt\,\int^\infty_{-\infty} dt'\, \frac{M^2_*}{r_t^3 {r_t'}^3} \cos \omega_{20}(t-t') \,.
\end{align}

\end{widetext}
 
 If we define (similar to \cite{press1977formation})
 \begin{align}\label{eqi}
 I_{m}(y)  = \int^\infty_0  \, \frac{dx}{(1+x^2)^2} \cos {[\sqrt{2} y (x+x^3/3)+2 m \tan^{-1} x]}\,,
 \end{align}
 $\Delta J_{\rm mode,z}$ in Eq.~\eqref{eqdjhyper} can be rewritten as
\begin{widetext}
 \begin{align}\label{eqjmodez}
 \Delta J_{\rm mode,z} = (Q^{22}_\xi)^2 \frac{6 \pi}{5 \omega_{22}} \frac{M^2_*}{G(M+M_*)R^3_0}[I^2_{-2}(\varpi_{22})-I^2_{2}(\varpi_{22})]\,,
 \end{align}
\end{widetext}
  with the renormalized frequency $\varpi$ given by
  \begin{align}
  \varpi \equiv \left [\frac{R^3_0}{G(M+M_*)} \right ]^{1/2} \omega\,.
  \end{align}

Similarly, the energy change described by Eq.~\eqref{eqehyper} can be rewritten as
 \begin{align}\label{eqemodef}
 \Delta E_{\rm mode} & = (Q^{22}_\xi)^2 \frac{6 \pi}{5} \frac{M^2_*}{G(M+M_*)R^3_0}[I^2_{-2}(\varpi_{22})+I^2_{2}(\varpi_{22})]\,\nonumber \\
 &+(Q^{20}_\xi)^2 \frac{4 \pi}{5 } \frac{M^2_*}{G(M+M_*)R^3_0} I^2_{0}(\varpi_{20})\,.
   \end{align} 
 This is consistent with the result in  \cite{press1977formation} obtained using the frequency domain calculation.
 
\subsubsection{Bound eccentric orbits}

Although a highly eccentric bound orbit near its pericenter can be well approximated by a parabolic orbit, we still present the energy and angular momentum transfer due to mode excitations explicitly, which is relevant for an evolving sequence of encounters.

A bound ($e<1$) eccentric orbit can be described by
\begin{align}
 r_t=& \frac{p_0}{1+e \cos \Phi} = a_0 (1-e \cos \tau)\,,\nonumber \\
 t= &\left [ \frac{a^3_0}{G(M_*+M)}\right ]^{1/2}(\tau - e \sin \tau)\,,\nonumber \\
 \tau=& \arccos \frac{e+\cos \Phi}{1+e \cos \Phi}\,,
 \end{align}
with $p_0$ being the semilatus rectum, $a_0$ the length of the semi-major axis, 
$e$ the eccentricity ($1-e \ll 1$ for highly eccentric orbits),
$\tau$ the mean anomaly, and $\Phi$ the true anomaly.
The pericenter distance $R_0$ equals $a_0(1-e)$. Following the same procedure as in Sec.~\ref{sec:hyper}, but with a new quantity $P_m$ defined to replace $I_m$:
 \begin{align}
 P_{m}(y)  =& \int^\pi_0  \, \frac{dx}{(1-e \cos x)^2} \nonumber \\
& \times \cos {\left [ y \frac{x -e \sin x}{(1-e)^{3/2}}+ m \cos^{-1} \frac{\cos x-e}{1-e \cos x} \right ] }\,,
 \end{align}
the angular momentum change of the reference star after each passage is

\begin{widetext}
\begin{align}\label{eqjinitd}
 \Delta J_{\rm mode,z} &= |Q^{22}_\xi|^2 \frac{3 \pi}{5 \omega_{22}} \frac{M^2_*}{G(M+M_*)a^3_0}[P^2_{-2}(\varpi_{22})-P^2_{2}(\varpi_{22})]
+ \frac{2 \sqrt{6 \pi/5} M_*}{a^{3/2}_0 [G(M+M_*)]^{1/2}} [P_{-2}(\varpi_{22})+P_{2}(\varpi_{22})] {\rm Im}[A^{(22)}_{\rm init} Q^{22}_\xi] \nonumber \\
&+ \frac{2 \sqrt{6 \pi/5} M_*}{\omega_{22} a^{3/2}_0 [G(M+M_*)]^{1/2}} [P_{-2}(\varpi_{22})-P_{2}(\varpi_{22})] {\rm Re}[\dot{A}^{(22)}_{\rm init} Q^{22}_\xi] \,,
   \end{align}
 where we no longer ignore the residual oscillation from previous encounters during a series of pericenter passages. Similarly the modal energy change is
 \begin{align}\label{eqeinitd}
  \Delta E_{\rm mode} & = |Q^{22}_\xi|^2 \frac{3 \pi}{5} \frac{M^2_*}{G(M+M_*)a^3_0}[P^2_{-2}(\varpi_{22})+P^2_{2}(\varpi_{22})]
 +|Q^{20}_\xi|^2 \frac{2 \pi}{5 } \frac{M^2_*}{G(M+M_*)a^3_0} P^2_{0}(\varpi_{20})\, \nonumber \\
 &+ \frac{ \omega_{22} \sqrt{6 \pi/5} M_*}{ a^{3/2}_0 [G(M+M_*)]^{1/2}} [P_{-2}(\varpi_{22})-P_{2}(\varpi_{22})] {\rm Im}[A^{(22)}_{\rm init} Q^{22}_\xi] \nonumber \\
 & +\frac{ \sqrt{6 \pi/5} M_*}{ a^{3/2}_0 [G(M+M_*)]^{1/2}} [P_{-2}(\varpi_{22})+P_{2}(\varpi_{22})] {\rm Re}[\dot{A}^{(22)}_{\rm init} Q^{22}_\xi] 
 - \frac{2\sqrt{\pi/5} M_*}{a^{3/2}_0 [G(M+M_*)]^{1/2}}\dot{A}^{(20)}_{\rm init} Q^{20}_\xi P_0(\varpi_{20})\,.
      \end{align}

\end{widetext}

\subsection{Mode damping}

Soon after each periastron passage, the distance between the stars grows
large enough so that the tidal interaction is no longer important until
the next periastron passage. Therefore
we shall approximate the mode evolution between two periastron
passages as free evolution damped by GW radiation. The
mode damping due to viscosity is expected to have a timescale longer
than the timescales such systems would remain in the aLIGO band (or
future ground-based detectors such as Einstein Telescope and Cosmic
Explorer), and shall be neglected in this analysis. 

The mode damping due to GW radiation can be approximated using the quadrupole formula:
\begin{align}\label{eqdampm}
\frac{d E_{\rm mode}}{d t} = -\frac{1}{5} \left \langle \frac{\partial^3 \mathcal{I}_{jk}}{\partial t^3} \frac{\partial^3 \mathcal{I}_{jk}}{\partial t^3} \right \rangle_t \,,
\end{align}
where $\langle \rangle_t$ is the temporal average (over timescales longer than the oscillation period but shorter than the decay time) and the trace-free quadrupole moment is 
\begin{align}\label{quadmom}
\mathcal{I}_{jk} = \int d^3 x \rho\, \left (x_j x_k - \frac{\delta_{jk}}{3} r^2 \right )\,.
\end{align}
Because the fluid motion following the f-mode eigenfunction   is trace-preserving for the quadrupole moment, the $r^2$ term in Eq.~\eqref{quadmom} can be dropped, so we have 
\begin{align}
\dddot{\mathcal{I}}_{jk} \approx \int d^3x \rho\, (x_j \dddot{\xi}_k+x_k \dddot{\xi}_i)\,,
\end{align}
where the $\mathcal{O}(\xi^2)$ terms  are neglected.
We shall rewrite Eq.~\eqref{eqdampm} as 
\begin{align}\label{eqerate}
\frac{d E_{\rm mode}}{d t} \approx -\frac{4}{5} \sum_n \langle \dddot{\mathcal{O}}^{(n)}_{\langle ab \rangle}  \dddot{\mathcal{O}}^{(n)}_{\langle ab \rangle} |A^{(n)}|^2 \rangle_t \,.
\end{align} 
During this ``free" (without tidal driving) evolution phase, the energy damping rate is proportional to the mode amplitude squared, and the mode energy is also proportional to its amplitude squared. As a result, the mode evolution can be modeled as a decaying oscillation with a fixed quality factor $Q_{lm}$ or decay rate $\gamma_{lm}=\pi f_{\rm mode}/Q_{lm}$. If a family of modes is initially excited, they should decay independently following their own decay rates because modes with different $l, m$ do not overlap in angular directions (the angular average is zero), and modes with the same $l, m$ but different overtones do not overlap in time (the temporal average is zero). For our purpose we mainly focus on $\ell=2$ modes, as they are the dominant modes excited after the periastron passage.

Consider a decaying oscillation of the (2,2) mode
\begin{align}
A^{22}(t) = \mathcal{A}^{22}(t) e^{- i \omega_{22} t} = \mathcal{A}^{22}_0 e^{-i \omega_{22} t-\gamma_{22} t} e^{i \phi_0}\,.
\end{align}

The mode energy, according to Eq.~\eqref{eqem} (assuming $Q_{22} \gg 1$), is
\begin{align}\label{eq22e}
E_{\rm mode} = \omega^2_{22} [\mathcal{A}^{22}(t)]^2\,.
\end{align}
The right hand side of Eq.~\eqref{eqerate} is given by
\begin{align}\label{eq22erate}
 -\frac{4}{5} \langle \dddot{\mathcal{O}}_{\langle ab \rangle}  \dddot{\mathcal{O}}_{\langle ab \rangle} \rangle_t = -\frac{1}{5} [Q^{(22)}_\xi]^2\times \frac{8 \pi}{15}  \omega^6_{22} [\mathcal{A}^{22}(t)]^2\,.
   \end{align}

Combining Eq.~\eqref{eq22erate} and Eq.~\eqref{eq22e}, we deduce the decay rate of the (2,2) mode to be
\begin{align}
\gamma_{22} = \frac{4 \pi}{75}  [Q^{(22)}_\xi]^2 \omega_{22}^4\,,
\end{align}
and correspondingly the quality factor is
\begin{align}\label{eqquality}
Q_{22} = \frac{75}{8 \pi}  [Q^{(22)}_\xi]^{-2} \omega_{22}^{-3}\,.
\end{align}

\section{Parabolic orbits of polytropic stars }\label{sec3}

In this section, we apply the formalism presented in Sec.~\ref{sec2} to
polytropic stars undergoing parabolic encounters,  which allows us to compare the results
with the full numerical solutions described in Sec.~\ref{sec4}. We find that the
total mode energy deposited onto stars can be comparable to the  energy radiated
by GWs ($\Delta E_{GW}$) during a close encounter. 
Given the various approximations employed in the analytic treatment, we do not
expect exact agreement. Better agreement may be achieved by incorporating PN treatment of the stars' oscillation and higher-order PN description of the orbit.

\subsection{Prediction for polytropic EOS}

To enable a simple comparison between the formalism in Sec.~\ref{sec2} and numerical simulations, we assume an equal-mass NS binary $M=M_*$, and a polytropic NS EOS, specifically $P =K \rho^{\Gamma}$ with $\Gamma=2$. For such stars, the  spherically symmetric equilibrium configuration can be obtained by solving
\begin{align}
& P'(r) =-\rho(r) U'(r) \nonumber \\
& U''(r)+\frac{2 U'(r)}{r} =4 \pi G \rho(r)\,,
\end{align}
with $U$ being the Newtonian potential.
 The corresponding solutions are
\begin{align}
& \rho(r) = \rho_c \frac{\sin (\pi r/R_*)}{\pi r/R_*} \,,\nonumber \\
& P(r) =K \left (\rho_c \frac{\sin (\pi r/R_*)}{\pi r/R_*} \right )^2 \,, \nonumber \\
& U(r) =-\frac{G M}{R_*} -4\pi G \rho_c R_*^2 \frac{\sin (\pi r/R_*)}{\pi r/R_*}\,,
\end{align}
with $K=2 G R_*^2/\pi$ and $M=4 \rho_c R_*^3/\pi$. Both $K$ and $\rho_c$ are fixed if we choose the NS radius for a given NS mass. According to the analysis in Appendix \ref{appendixA}, the normalized f-mode eigenfrequencies for such stars are given by 
\begin{align}\label{eqmodef}
\varpi_{20}=\varpi_{2,\pm2} = 0.8676 \left( \frac{R_0}{R_*}\right)^{3/2}\,.
\end{align}
The frequency degeneracy is due to the spherical symmetry of the
background solution, such that the mode frequency is independent of
the azimuthal wave number. Similarly, the mode overlap constants for
the $22$ and $20$ modes are the same, as
$\xi^{(n)}_R$ and $\xi^{(n)}_S$ are also independent of $m$ due to the
spherical symmetry (the $21$ mode is irrelevant for computing
energy and angular momentum transfer, according to
Eq.~\eqref{eqemodef} and Eq.~\eqref{eqjmodez}). 
The detailed form of the radial dependence of wave
functions can be obtained using the method described in Appendix
A. The overlap constants are
\begin{align}\label{eqoc}
Q^{22}_{\xi} = Q^{20}_\xi = 0.558 (M R^2_*)^{1/2}\,,
\end{align}
which implies that the quality factors (c.f. Eq.~\eqref{eqquality}) are
\begin{align}\label{eqqn}
Q_{2,\pm 2} =Q_{20}\approx 5.2 \left ( \frac{R_*}{M}\right )^{5/2}\,.
\end{align}

\begin{figure}[tb]
\includegraphics[width=8.4cm]{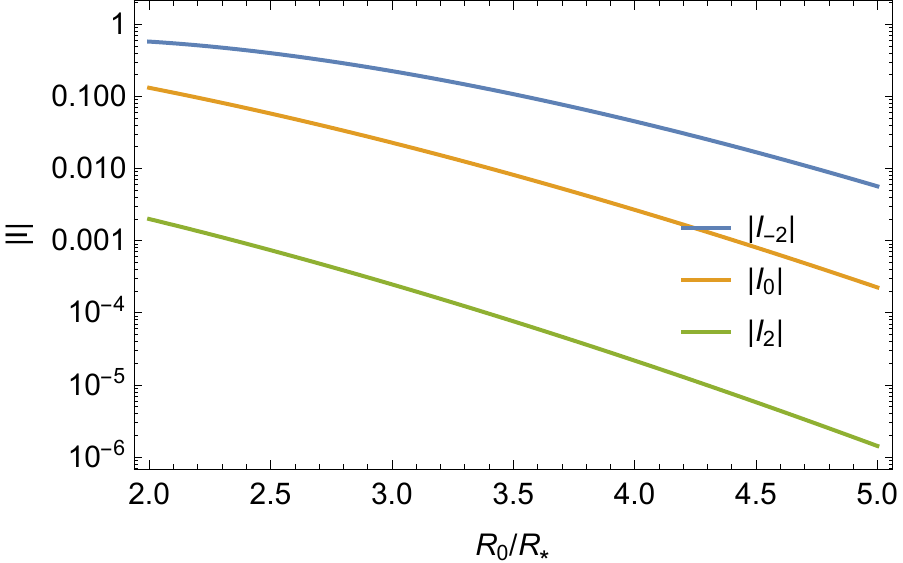}
\caption{$|I_{\pm 2}|$ and $|I_{0}|$ (Eq.~\eqref{eqi}) as  functions of $R_0/R_*$. In the
Newtonian limit, when $R_0 = 2 R_*$ the surfaces of two undeformed stars are in
marginal contact at the periastron passage. 
}
\label{fig:Iplot}
\end{figure}

From Eq.~\eqref{eqmodef} and Eq.~\eqref{eqi}, the mode excitation coefficients (as functions of $R_0/R_*$) are shown in Fig.~\ref{fig:Iplot}.
Notice under the convention  in Eq.~\eqref{eqi}, the magnitude of $I_{-2}$ is much greater than $I_{2}$, which has to do with the fact that the star's motion is counterclockwise as described by Eq.~\eqref{eqntrajectory}. As a result, when we use Eq.~\eqref{eqemodef} to compute mode energy, the $m=-2$ piece dominates over other parts. The study in \cite{chirenti2016gravitational} only includes the $m=2$ piece, which explains why the result therein is much smaller than the values inferred by numerical simulations.

With the excitation coefficients shown in Fig.~\ref{fig:Iplot}, and the overlap constants obtained from  Eq.~\eqref{eqoc}, we can write the mode energy in Eq.~\eqref{eqemodef} as
\begin{align}\label{eqemodefn}
 \Delta E_{\rm mode} & =  0.587\frac{G M^2 R^2_*}{R^3_0}[I^2_{-2}(\varpi_{22})+I^2_{2}(\varpi_{22})]\,\nonumber \\
 &+ 0.391\frac{G M^2 R_*^2}{R^3_0} I^2_{0}(\varpi_{20})\,.
   \end{align} 
and mode angular momentum
\begin{align}\label{eqjmodefn}
 \Delta J_{\rm mode} & =  0.478 \frac{G^{1/2} M^{3/2} R^{7/2}_*}{R^3_0}[I^2_{-2}(\varpi_{22})-I^2_{2}(\varpi_{22})]\,.
   \end{align}

Without considering mode excitations, the GW energy emitted during a periastron passage can be estimated as \cite{peters1964gravitational,chirenti2016gravitational}:
\begin{align}
\Delta E_{\rm GW} \approx 7 \pi \frac{(G M)^{7/2} M}{c^5 R_0^{7/2}}\,,
\end{align}   
and the angular momentum carried away by GWs is 
\begin{align}
\Delta J_{\rm GW} \approx 6 \pi \frac{(G M)^{3} M}{c^5 R_0^{2}}\,.
\end{align}  
 
 As a result, the ratio between the mode energy deposited into both stars and the energy carried away by GWs during the main burst is
 \begin{align}\label{eqeratio}
 \frac{2\Delta E_{\rm mode}}{\Delta E_{\rm GW}} \approx &\left ( \frac{G M}{c^2 R_*}\right )^{-5/2} \left ( \frac{R_0}{R_*}\right )^{1/2} \times 2\times 10^{-2} \times \nonumber \\
 &\left \{ 2.69 [  I^2_{-2}(\varpi_{22})+ I^2_{2}(\varpi_{22})]+1.78 I^2_{0}(\varpi_{20})\right \}\,.
 \end{align}
Similarly, the ratio between the mode angular momentum deposited into both stars and the angular momentum carried away by GWs during the main burst is
 \begin{align}\label{eqjratio}
 \frac{2\Delta J_{\rm mode}}{\Delta J_{\rm GW}} \approx &\left ( \frac{G M}{c^2 R_*}\right )^{-5/2} \left ( \frac{R_0}{R_*}\right )^{-1} \times 0.05 \times \nonumber \\
 &\left \{ [  I^2_{-2}(\varpi_{22})- I^2_{2}(\varpi_{22})])\right \}\,.
 \end{align}
 
\begin{figure}[tb]
\includegraphics[width=8.4cm]{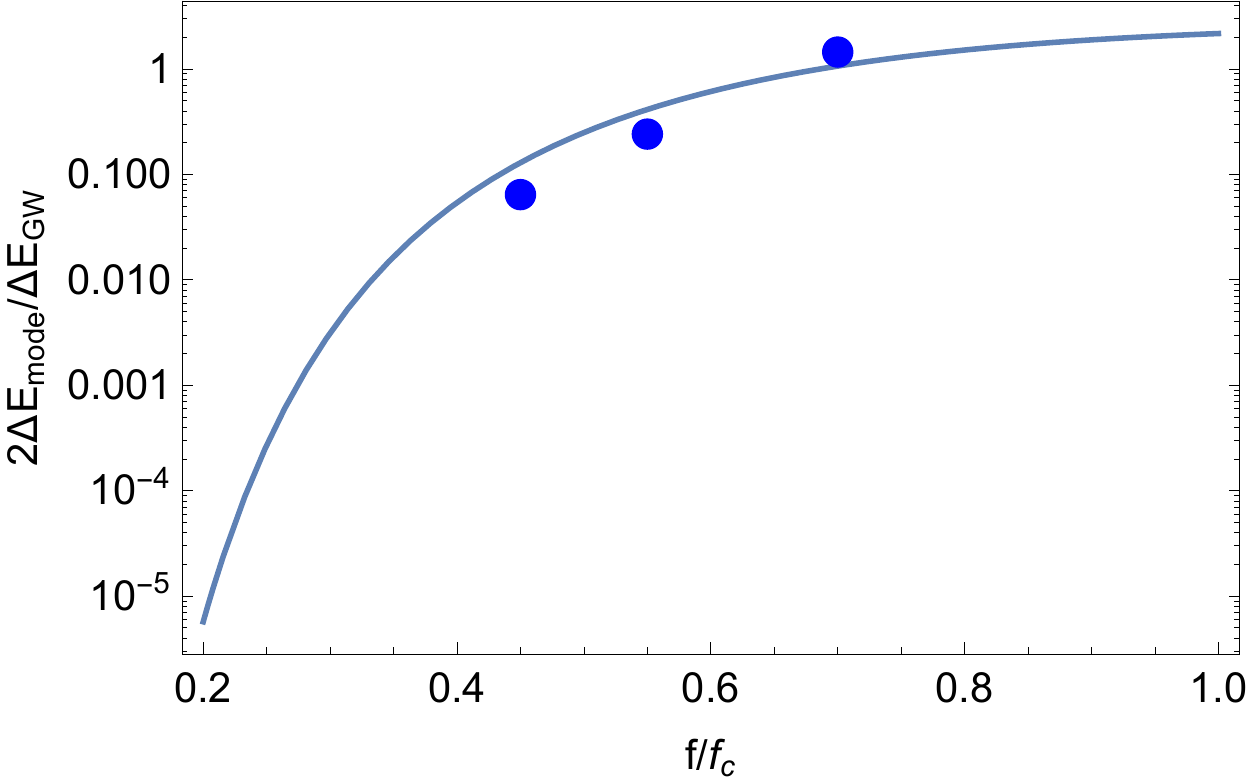}
\includegraphics[width=8.4cm]{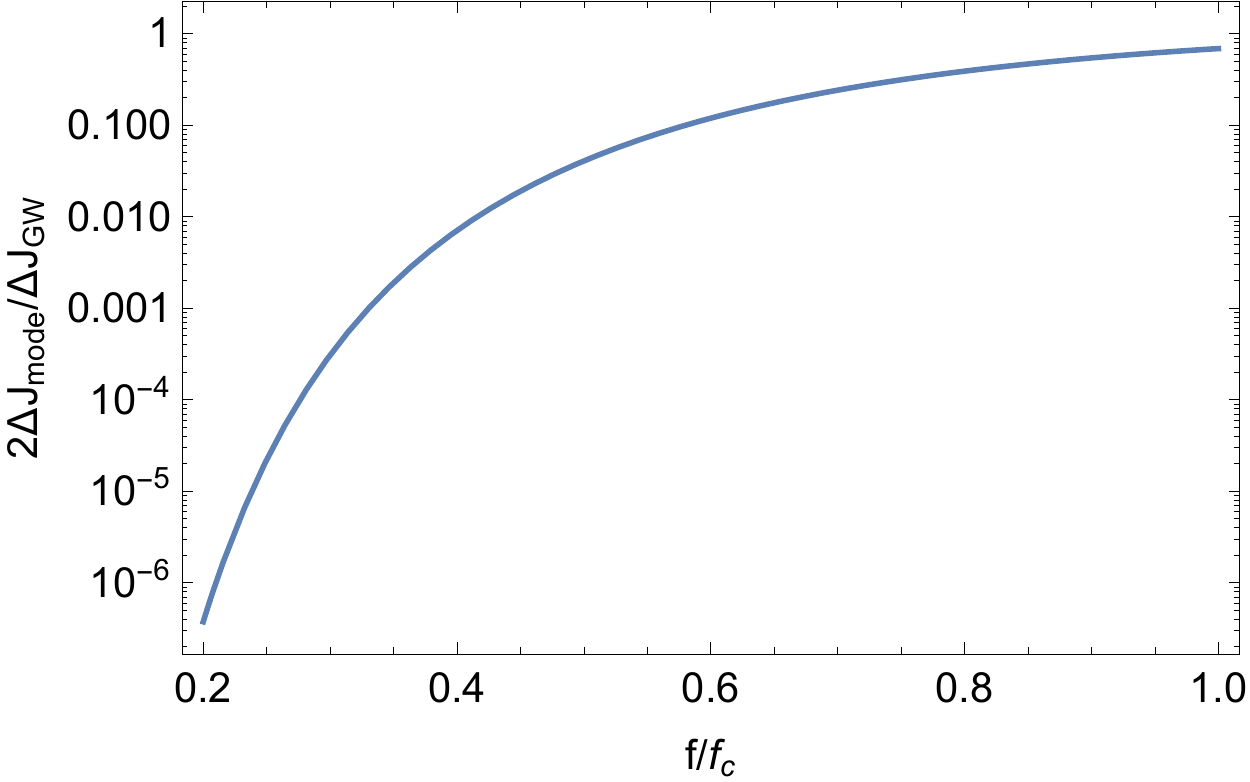}
\caption{The ratio between the total mode energy (angular momentum) deposited onto stars and the energy (angular momentum) carried away by GWs during the main burst, as a function of the normalized periastron frequency $f/f_c$. Here $R_* c^2/(G M)$ is taken to be $5.88$.
   The dots shown in the top plot represent estimates from the numerical simulations listed in Table.~\ref{tab:emode}.
}
\label{fig:ratioplot}
\end{figure}

The dominant contribution comes from the $m=-2$ mode with our
conventions. As a concrete example, take a NS with compaction $G
M/(c^2 R_*) \sim 0.17$ (e.g.  $M \approx 1.4$ $M_\odot$ and $R_*
\approx 12$ km). In Fig.~\ref{fig:ratioplot} we plot the ratio
described in Eq.~\eqref{eqeratio} and Eq.~\eqref{eqjratio} as a
function of the ``periastron frequency" $f$ (of the orbit, hence
half the GW frequency). This frequency is
introduced in \cite{loutrel2017eccentric,chirenti2016gravitational}
(also discussed in Sec.~\ref{sec4}), and represents the peak frequency
of the main burst. In the Newtonian limit, $f$ is proportional to
$R_0^{-1.5}$. In Fig.~\ref{fig:ratioplot}, $f_c := \sqrt{M/(2 R_*)^3}/\pi$ denotes the
periastron frequency for the closest possible passage without the
stars colliding.  For comparison, in this figure we also show results
from the full numerical solutions described in Sec.~\ref{sec4}. These
agree with the model to within a factor of two, even at the
relativistic velocities considered.  The plot shows that, for very
close pericenter passages, the mode energy/angular momentum deposited
into the stars can be of the same order of magnitude as the
energy/angular momentum carried away by GWs. This means that it is
crucial to include the mode dynamics in order to accurately model the
orbital evolution.

\section{Simulations in full GR}\label{sec4}

In order to validate the model described in the previous section, and determine
its accuracy into the relativistic regime, we simulate several cases consisting
of compact object binaries undergoing a close encounter using GR coupled to
hydrodynamics.  This work makes use of similar methods to those of previous
studies of eccentric binary
mergers~\cite{East:2011xa,East:2012ww,Paschalidis:2015mla,East:2015vix,East:2016zvv},
which we just briefly review here.

\subsection{Numerical Methods}
These simulations are carried out by solving the full Einstein equations
coupled to hydrodynamics using the methods described in~\cite{code_paper}. For
ease of comparison to the approximate model described in this work, we restrict
ourselves to a $\Gamma=2$ EOS. We
make use of adaptive mesh refinement with seven levels of refinement.
The base-level resolution has $201^3$ points, while the 
finest-level resolution has approximately $100$ points across the NS diameter.
For simulations with BHs, we use one additional mesh refinement level, which
gives a factor of two better resolution, around the BH.
For one of the NS-NS cases, we also perform a lower resolution simulation with
$0.6\times$ the above resolution, in order to estimate truncation error.

\subsection{Initial Data and Cases}
We study binary NS close encounters with several different impact
parameters, and then follow the oscillations in the stars during the
long outgoing part of the elliptic orbit.  For comparison, we also
consider an equal mass BH-NS case with a non-spinning BH and the same
orbital parameters as one of the binary NS cases. Initial data is
constructed using the methods described in~\cite{idsolve_paper}. We
choose the initial velocities and positions of the compact objects at
large separation ($d=50M_{\rm tot}$, where $M_{\rm tot}$ is the total
mass of the system) based on a marginally bound Newtonian orbit with a
specified periapse distance $R_0$.  The actual periapse distance of
the binary will be different (due to gauge effects, relativistic corrections, etc.), 
and we fix the parameters used
for comparing to the approximate model based on the frequency of the
fly-by gravitational waveform. This allows a largely gauge-invariant
comparison with the model we presented in the previous section.  We
consider several equal mass binary NS cases with $R_0/M_{\rm
  tot}=10.0$, 11.5, and 13. For the comparison to an equal mass BH-NS system, we 
use a single NS case with $R_0/M_{\rm tot}=11.5$.
In all cases we choose a NS with
$M_*/R_*=0.17$.

\subsection{Results}

\begin{figure}
    \begin{center}
        \includegraphics[width=\columnwidth,draft=false]{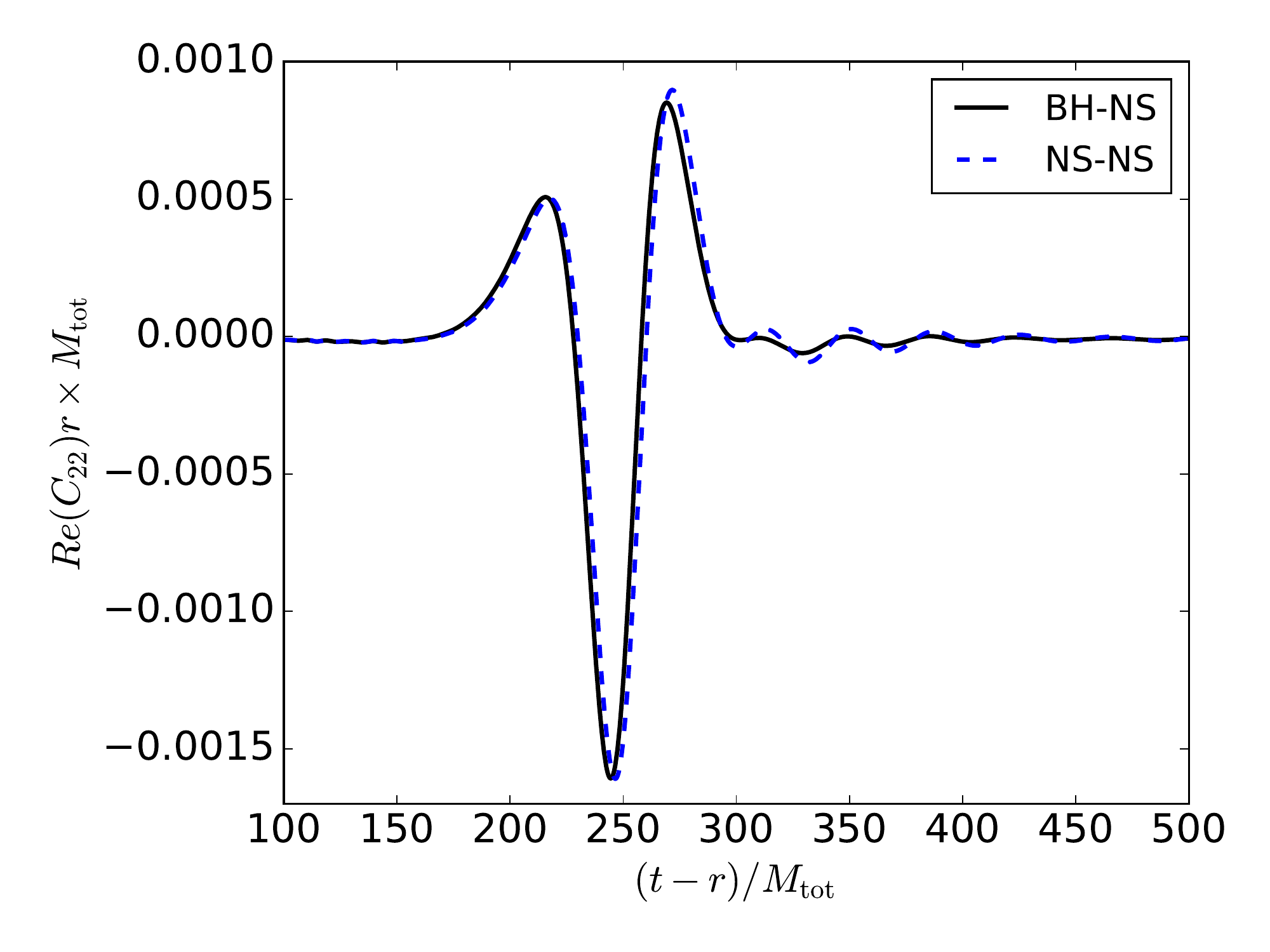}
        \includegraphics[width=\columnwidth,draft=false]{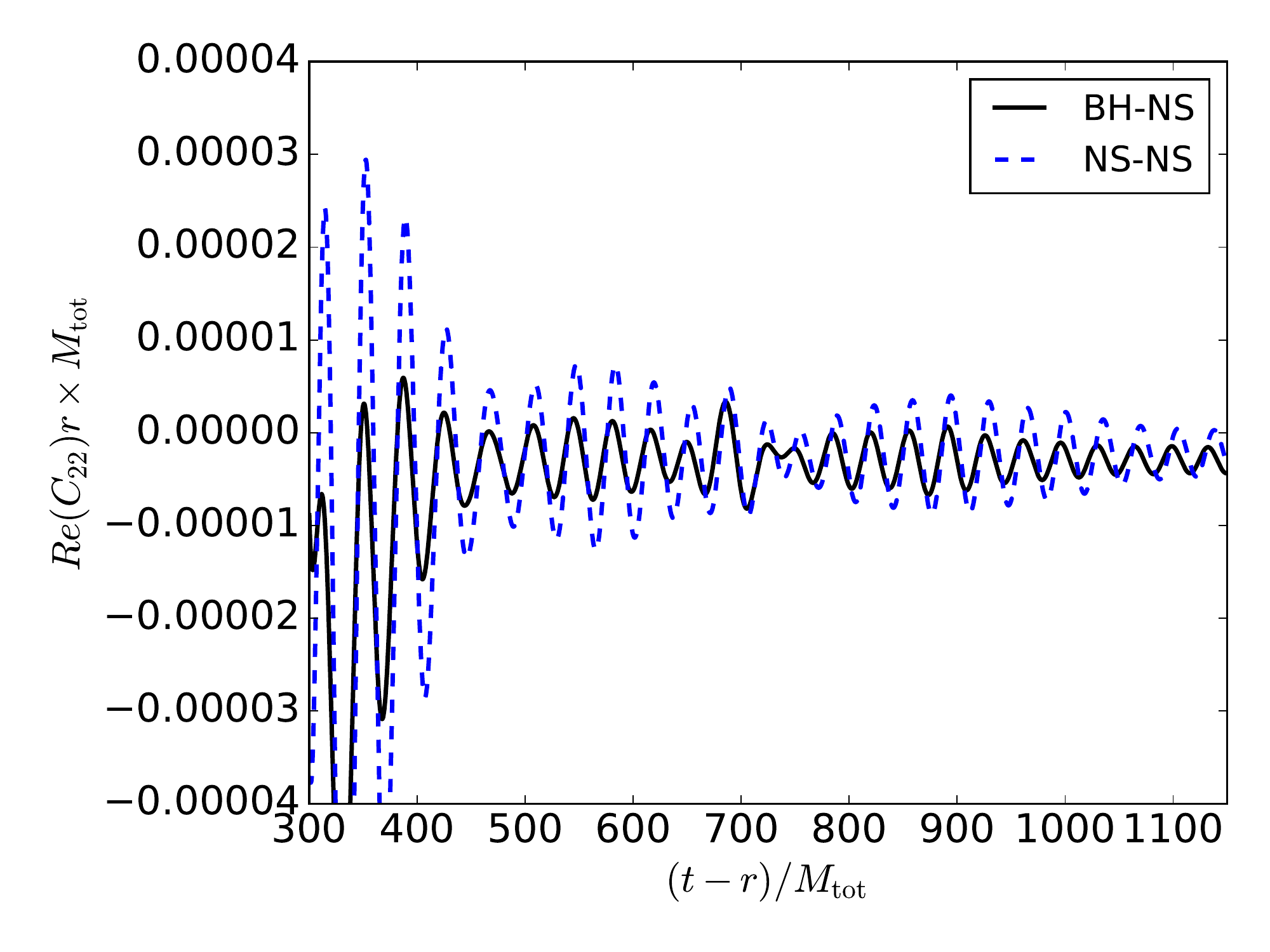}
        \includegraphics[width=\columnwidth,draft=false]{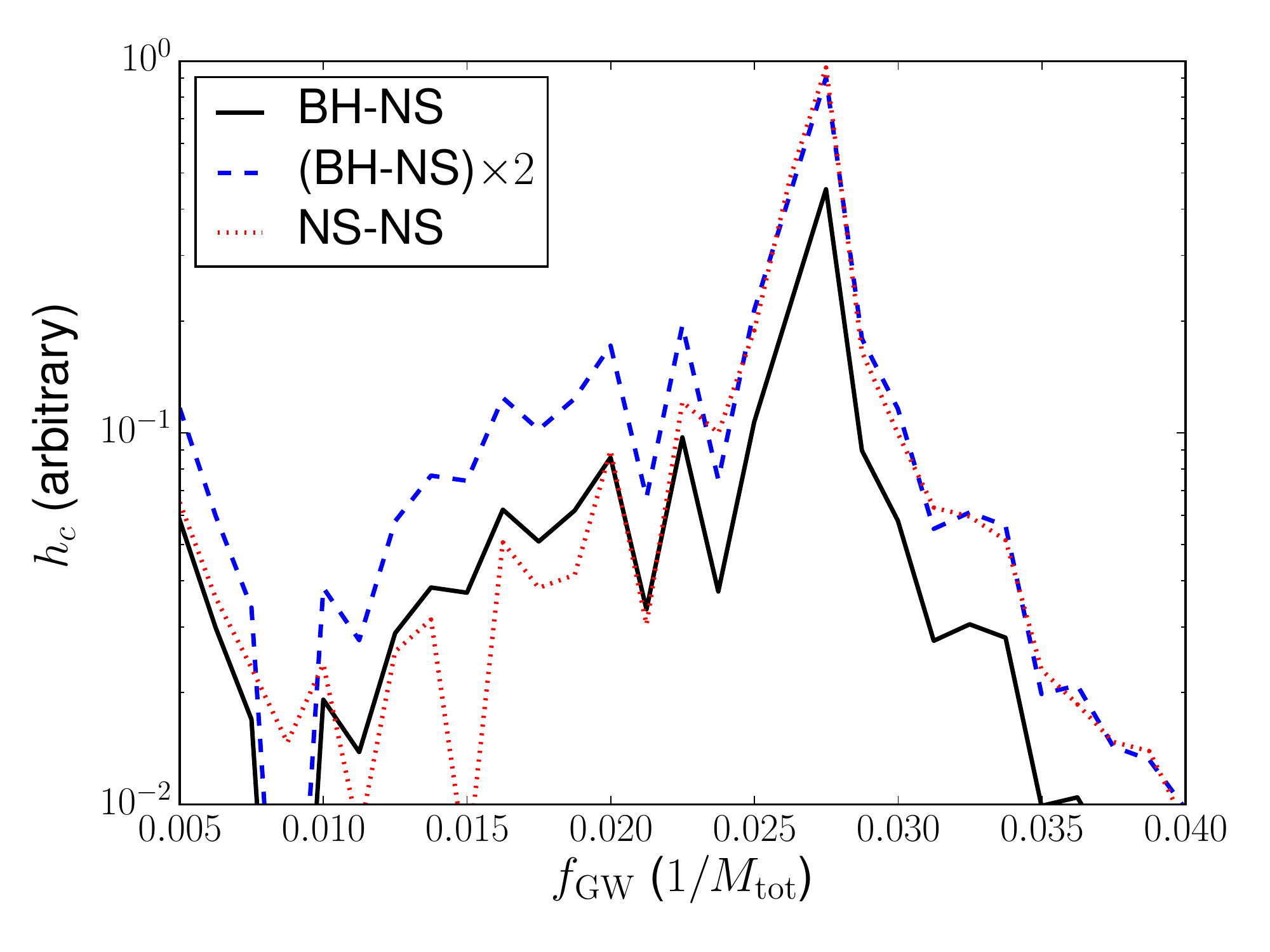}
    \end{center}
    \caption{ A comparison of the GWs from a BH-NS and NS-NS case with
      $R_0/M_{\rm tot}=11.5$.  The top and middle panels show the real
      part of the $(2,2)$ component of $\Psi_4$, the former
      emphasizing the GW burst from the close encounter, and the
      latter emphasizing the GW oscillations from the $f$-mode
      excitations in the NS(s). The bottom panel shows the post fly-by
      GWs ($t-r>500M_{\rm tot}$) in the frequency domain. The peak at
      approximately $f_{\rm GW} \approx0.027/M_{\rm tot}$ associated with the
      $f$-mode oscillations is a factor of two larger in the binary NS
      case, as expected if the NSs are tidally perturbed by the same
      amount in both cases.
        \label{fig:bhns_nsns_gw}
    }
\end{figure}

In Fig.~\ref{fig:bhns_nsns_gw} we show a comparison of the GWs from a
NS-NS and BH-NS case with the same orbital parameters. The fly-by part
of the waveform matches well between the two cases, indicating that the
incoming orbits are very close.  The peak instantaneous GW frequency,
calculated from the time derivative of the phase of the $(2,2)$
component of $\Psi_4$ is $\omega M_{\rm tot}=0.11$, and differs by
$<1\%$ in the two cases.  This corresponds to normalized periastron
frequency of $f/f_c=0.55$. The amount
of energy radiated in GWs around the periapse passage is $\Delta
E_{\rm GW}/M_{\rm tot}\approx 6.4\times10^{-4}$.

After the fly-by there are high frequency GW oscillations in both cases. By
going to the frequency domain we can see these GW oscillations are primarily at
the expected $f$-mode frequency, and that the amplitude in the binary NS case
is twice that of the BH-NS case, indicating that to a good approximation the
stars are tidally perturbed by the same amount. In the middle panel of
Fig.~\ref{fig:bhns_nsns_gw}, one can see a lower frequency modulation in the GW
signal.  The fact that this occurs in both the binary NS and BH-NS cases
indicates that it is not due to interference effects, and is perhaps instead
due to the presence of more than one fluid oscillation mode.

\begin{figure}
    \begin{center}
        \includegraphics[width=\columnwidth,draft=false]{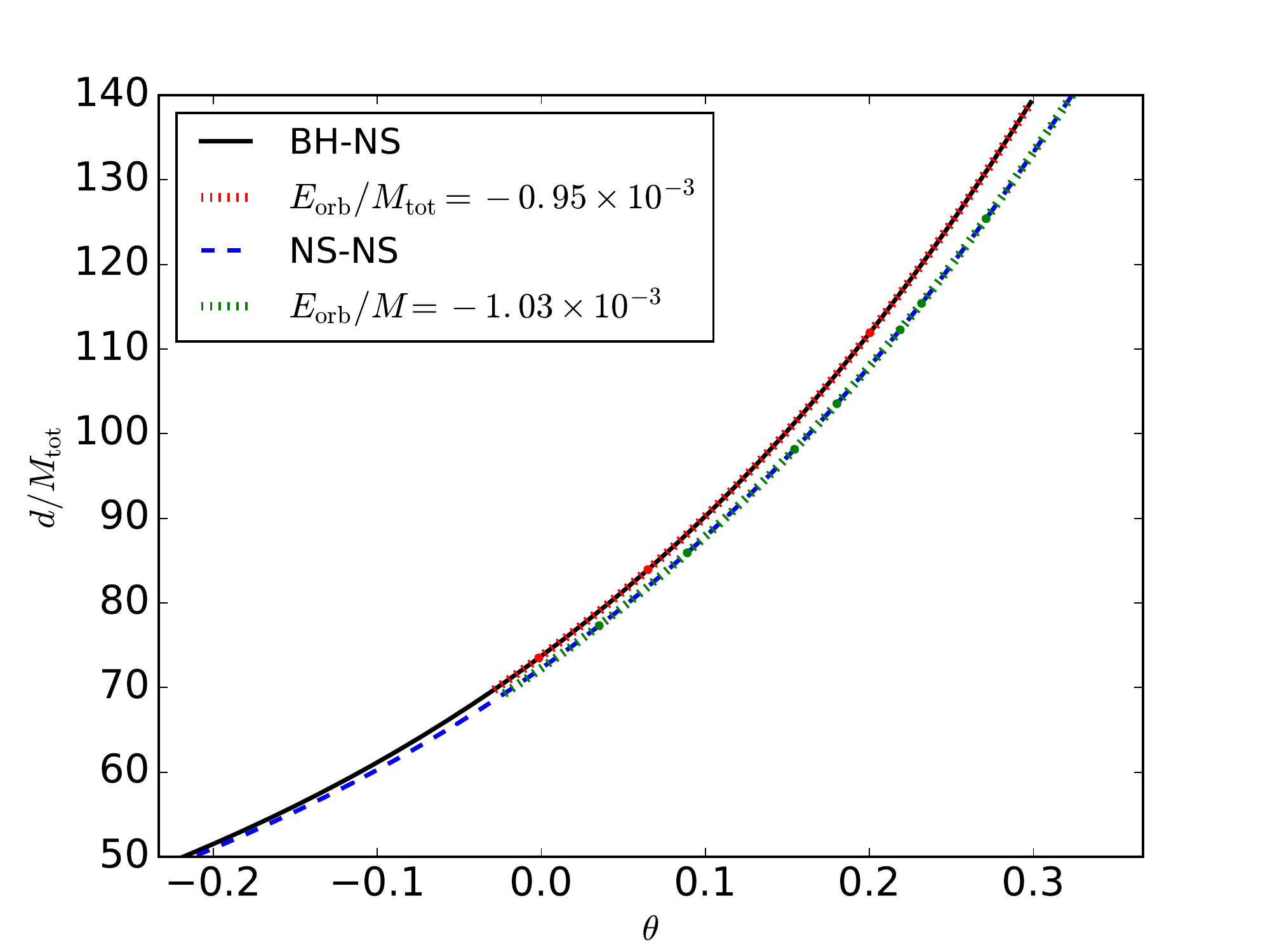}
    \end{center}
    \caption{
        The coordinate separation versus orbital angle at large distances
        post-flyby for a BH-NS and NS-NS system with the same initial 
        binary parameters. We also show the Newtonian
        orbits with the best-fit parameters for these two cases.
        \label{fig:bhns_nsns_orbit}
    }
\end{figure}

We can estimate the amount of energy lost to tidal excitations by comparing the
orbits in the two cases. In Fig.~\ref{fig:bhns_nsns_orbit} we show the
coordinate separation of the binary as a function of angle, at large values
post-flyby.  Fitting this to a Newtonian orbit, and taking the difference
of the resulting values for the orbital energy for the BH-NS and NS-NS cases
and attributing it to the energy of the tidal excitations in a single star
gives $\Delta E_{\rm mode}/\Delta E_{\rm GW}=0.13$.\footnote{
Our method of constructing initial data and resolving the constraints causes
the orbital energy of the binary to be slightly negative.  
If instead of using the difference between the BH-NS and the NS-NS cases, one just used the orbital energy
estimate for the NS-NS case, and assumed it was initially zero before the close
encounter, we would have obtained a somewhat larger value of $\Delta E_{\rm mode}/\Delta E_{\rm GW}=0.3$. 
}
In comparison, the analytic model predicts $\Delta E_{\rm mode}/\Delta E_{\rm GW}=0.22$.

We use the above estimate of $\Delta E_{\rm mode}$ as a reference
value, and also consider how the tidal excitation compares as a
function of impact parameter ---or equivalently, periastron
frequency---for a higher and lower case. The results are summarized in
Table~\ref{tab:emode}. All cases show the expected peak in the
characteristic strain at the f-mode frequency as in the bottom panel
of Fig.~\ref{fig:bhns_nsns_gw}, and here we have estimated $\Delta
E_{\rm mode}$ by assuming that it is proportional to the strain within
this peak (within $10\%$ of the f-mode frequency), and using the above
comparison to the BH-NS to fix the overall magnitude. We also note
that by comparing two different resolutions for the largest impact
parameter (lowest value of $f/f_c$ case), and assuming second order
convergence~\footnote{Our code is second-order convergent when shocks
  are absent.}, we estimate that the truncation error in the
calculation of $\Delta E_{\rm mode}$ is $\sim20\%$ (and it is an
underestimated), while the truncation errors in $f/f_c$ and $E_{\rm
  GW}$ are smaller, $\lesssim1\%$. Likely more significant are
systematic effects in estimating orbital energy based on comparing
coordinate trajectories for the BH-NS and NS-NS cases, and measuring
tidal excitations based on the GW emission. Since our main purpose
here is to establish the rough accuracy of the analytic model for
tidal excitations, we leave a more in-depth study of this to future
work.

\begin{table}[h!]
  \centering
  \caption{ 
    The first two columns are the characteristic frequency and energy emitted in GWs during pericenter
    passage. The last two columns are an estimate of the amount of energy deposited in tidal excitations
    (per star) from three equal-mass, nearly parabolic binary NS encounters from the
    full GR calculation and the analytic model.
    }
  \label{tab:emode}
  \begin{tabular}{c|c|c|c}
    \hline
      $f/f_c$ & $\Delta E_{\rm GW}/M_{\rm tot}$ & $\Delta E_{\rm mode}/\Delta E_{\rm GW}$ (sims.) & 
      (model)\\
    \hline
      0.70 & $15.0\times10^{-4}$& 0.72 & 0.53 \\
    0.55 & $6.4\times10^{-4}$   & 0.12   & 0.22 \\
    0.45 & $3.4\times10^{-4}$   & 0.032   & 0.06 \\
  \end{tabular}
\end{table}

\section{Trajectory model}\label{sec5}
A trajectory and waveform model for highly eccentric, binary BH systems was
developed in \cite{loutrel2017eccentric}. The key idea is to divide the whole
trajectory into a series of PN elliptical orbits attached by a change of orbital
parameters that occurs during the pericenter
passage process. The shift of orbital eccentricity and energy/angular
momentum can be obtained by computing the accumulated GW flux within each cycle. 

Based on a PN description of the orbital evolution, the BH waveform model then
contains four physical quantities: $t_i, \delta t_i, f_i, \delta f_i$ for the
$i$th cycle. Here $t_i$ is the coordinate time for the $i$th periastron
passage while $f_i$ is the ``periastron frequency", defined as
\cite{loutrel2017eccentric}
\begin{align}
f_i  \equiv \frac{1}{2 \pi \tau_{\rm GW}} \equiv \frac{1}{2 \pi} \frac{\rm pericenter \,\,\,velocity}{\rm pericenter \,\,\, distance}\,,
\end{align}
which reduces to
\begin{align}
f_i =\frac{1}{2 \pi} \frac{[(M+M_*)(1+e)]^{1/2}}{R_0^{3/2}}
\end{align}
in the Newtonian limit. 

Phenomenologically, $2 f_i$ well approximates the central
burst frequency of GWs corresponding to the $i$th periastron passage. With the
predictions of $t_i$ and $f_i$ for the central arriving time and frequency of each
GW burst, one can extract a sequence of centroids of size $\delta t_i \times
\delta f_i$ in the time-frequency diagram around $(t_i, f_i)$, compute the
spectrum within each centroid, and possibly stack the spectra power of
different centroids to boost the total signal-to-noise ratio
\cite{East:2012xq,huerta2014accurate}.  The choices for $\delta t_i$ and $\delta f_i$ are
rather flexible; in \cite{loutrel2017eccentric} they are chosen to be
constant multiples of $\tau_{\rm GW}$ and $f_i$.

We notice that if the orbital evolution and GW emission can be modeled with
enough accuracy, such that the phase error of the waveform for the whole
duration is controlled within $\mathcal{O}(1/{\rm SNR})$ (with ${\rm SNR}$
being the signal-to-noise ratio of a typical event), it is preferable to use a
single waveform covering the whole duration for both detection and parameter
estimation purpose. This requirement is challenging because the binary dynamics
could be quite nonlinear near pericenter passages, and the total duration
could contain many cycles. It is computationally prohibitive to perform
numerical simulations of this type to calibrate the theoretical waveforms
designed for such systems.  Nevertheless, one possibility is to use  a waveform
calibrated by numerical relativity when the NSs are close to each other, and then
match it to a PN waveform when the two stars are far apart.  We shall leave
further investigation of this to future work.

\subsection{Order of magnitude estimates}

As discussed in Sec.~\ref{sec3}, and explicitly shown in
Fig.~\ref{fig:ratioplot}, the energy/angular momentum deposited into the stars
may be of the same order of magnitude as the energy/angular momentum carried
away by GWs during close pericenter passages. In this case it is necessary to
include the effect of mode excitations into the evolution model for such
orbits.

To quantify the regime of importance for mode excitations, we notice that $2
\Delta E_{\rm mode}/\Delta E_{GW}$ in Fig.~\ref{fig:ratioplot} exceeds $1\%$
only when $f/f_c \ge 1/3$, which means that 
\begin{align} 
    f \ge \frac{f_c}{3} \sim
    3.9\times 10^{2} {\rm \ Hz} \left ( \frac{M}{1.4\ M_\odot}\right )^{1/2} \left
    ( \frac{R_*}{12 {\rm \ km}}\right )^{-3/2}\,.  
\end{align} 

Another question is whether it is necessary to include the oscillations
inherited from the previous pericenter passage as in Eq.~\eqref{eqeinitd} and
Eq.~\eqref{eqjinitd}, instead of setting $A_{\rm init}$ and $\dot{A}_{\rm init}$ to
be zero before each pericenter passage. To answer this question, we compare the
mode decaying timescale $1/\gamma_{22}$ with the orbital period (in the
Newtonian limit)
\begin{align}
T & = 2 \pi \left [\frac{G (M+M_*)}{a_0^3} \right ]^{-1/2} \nonumber \\
&= 2 \pi \left [ \frac{G (M+M_*)}{R_0^3} \right ]^{-1/2} (1-e)^{-3/2}\,.
\end{align}
We conclude that the initial oscillations are negligible if (assuming $M=M_*$)
\begin{align}\label{eqec}
1-e \ll 0.63 \frac{R_0}{R_*} \left ( \frac{G M}{R_* c^2}\right )^{5/3}\,.
\end{align}

If we consider the type of NS assumed to produce Fig.~\ref{fig:ratioplot} and
consider the closest passage with $R_0 = 2 R_*$, the above inequality can be
translated to $1-e \ll 0.07$.  For such binaries (with Eq.~\eqref{eqec}
satisfied), it is no longer necessary to track the evolution of modes, as the
lifetime of the modes is smaller than the time between pericenter
passages. However, in order to compute the energy and angular momentum change
of the orbit after pericenter passages, it is still necessary to include the
mode contributions using Eq.~\eqref{eqeinitd} and Eq.~\eqref{eqjinitd}, with
$A_{\rm init}$ and $\dot{A}_{\rm init}$ set to zero.

In the special case studied in Sec.~\ref{sec4}, the two NSs are assumed to be
identical, which means that their f-modes have the same frequency. The GWs
generated by f-mode oscillations from two stars may beat with each other,
depending on the separation of two stars and the sky direction of observers.
This beating starts to become important if the orbital separation is comparable to,
or larger than, the half-wavelength of the f-mode GWs:
\begin{align}
1\le \frac{a_0(1+e)}{\lambda/2} \approx \frac{2 R_0 \omega_{22}}{\pi (1-e) c}\,,
\end{align}
or equivalently
\begin{align}
1-e \le  0.78 \frac{R_0}{R_*} \left ( \frac{G M}{R_* c^2}\right )^{1/2}\,.
\end{align}

In reality, the masses of individual NSs within the binary will be different
(we denote the mass difference as $\Delta M=M-M_*$). According to the f-mode
frequency formula (see Eq.~\eqref{eqmodef}), we have $\delta f/ f \sim 0.5
\Delta M/M$. The beating generally loses constructive interference after $f/2\delta f \sim 
M/\Delta M$ oscillation cycles. 

\subsection{Trajectory model}

The quasi-Keplerian (QK) description of eccentric orbits of a compact binary is
well explained in \cite{blanchet2014gravitational}. In this framework, the 1PN
motion of the binary in the radial, time, and angular directions can be written in
the following form:
\begin{align}\label{eqtra1pn}
r &= a_r (1- e_r \cos u)\,\nonumber \\
\ell &= u-e_t \sin u +\mathcal{O}(c^{-4})\,,\nonumber \\
\frac{\phi-\phi_p}{K} & = v+\mathcal{O}(c^{-4}) = 2 \arctan \left [ \left ( \frac{1+e_\phi}{1-e_\phi}\right )^{1/2} \tan \frac{u}{2}\right ]\,,
\end{align}
where $v$ is the true anomaly, $u$ is the eccentric anomaly, $\ell = n (t-t_p)$ is the mean anomaly
with $n=2\pi/T $ being the mean motion, and $P$ is the period. In addition, $K$
is related to the periastron precession angle  per cycle $\Delta \phi_p$ by
$\Delta \phi_p =2 \pi (K-1)$. The above QK representation has been generalized
to 2 PN \cite{damour1988higher,schafer1993second,wex1995second} and 3 PN orders
\cite{memmesheimer2004third}. The orbital eccentricities $e_t$, $e_r$, and $e_\phi$, and
additional parameters $n$, $K$, and $a_r$ are all functions of the orbital energy and
angular momentum (see Eq. (345) in \cite{blanchet2014gravitational}), and we
introduce a new set of parameters to match the convention in
\cite{blanchet2014gravitational}:
\begin{align}\label{eqrej}
\epsilon = -\frac{2 E}{\mu c^2},\quad j = -\frac{2 E h^2}{\mu^3},
\end{align}
with $\mu = (M+M_*) \nu $, $\nu = M M_*/(M+M_*)^2$ and $h = J/G(M+M_*)$. In order to obtain a mapping for orbital parameters from one cycle to the next, we first compute the energy and angular momentum change within the current cycle:
\begin{align}\label{eqEJmap}
E_{i+1} & = E_i +\Delta E_{\rm GW,i}+\Delta E_{\rm mode,i}\,\nonumber \\
J_{i+1} & = J_i +\Delta J_{\rm GW,i}+ \Delta J_{\rm mode,i}\,.
\end{align}

The energy and angular momentum radiated by the orbital motion within the $i$th cycle are
\begin{align}\label{eqDEJmap}
\Delta E_{\rm GW,i}  & = \langle \dot{E}_{\rm GW,i}\rangle T_i = \langle \dot{E}_{\rm GW,i}\rangle \frac{2 \pi}{n_i}\,,\nonumber \\
\Delta J_{\rm GW,i}  & =  \langle \dot{J}_{\rm GW,i}\rangle \frac{2 \pi}{n_i}
\end{align} 
where $n_i$ is given by Eq. ($347a$) in \cite{blanchet2014gravitational}. The 3PN orbital-averaged energy flux can be obtained from Eq. ($355$) in \cite{blanchet2014gravitational},
and the 3PN averaged angular momentum can be found in \cite{arun2009third}. For the mode energy and angular momentum change we use Eq.~\eqref{eqeinitd}, Eq.~\eqref{eqjinitd} and Eq.~\eqref{eqrej}. There is however one subtlety, which is to include the residual oscillation from the previous pericenter passage into the calculation, i.e., the terms involving $A_{\rm init}$ and $\dot{A}_{\rm init}$.
Based on Eq.~\eqref{eqamap}, we denote $A_{\rm init,i}=A_{\rm init}(t_i),\, \dot{A}_{\rm init,i} = \dot{A}_{\rm init}(t_i)$, and explicitly write down their mapping relations as (with $Q \gg 1$)
\begin{widetext}
\begin{align}\label{eqa2map}
A^{22}_{\rm init, i+1} & = \left [A^{22}_{\rm init,i} +i \sqrt{3 \pi/10} \frac{M_*}{n a^3_r}Q_\xi (P_{-2}-P_2) \right]\left [ \cos (\omega T_i) +\frac{\gamma}{\omega} \sin (\omega T_i)\right ] e^{- \gamma T_i}\nonumber \\
&+ \left [\sqrt{3 \pi/10} \frac{M_*}{n a^3_r}Q_\xi (P_{-2}+P_2)+\dot{A}^{22}_{\rm init,i } \right ] \frac{ \sin (\omega T_i)}{\omega} e^{- \gamma T_i}\nonumber \\
\dot{A}^{22}_{\rm init, i+1} & =\left [\sqrt{3 \pi/10} \frac{M_*}{n a^3_r}Q_\xi (P_{-2}+P_2)+\dot{A}^{22}_{\rm init,i } \right ]  \left [ \cos (\omega T_i) -\frac{\gamma}{\omega}\sin (\omega T_i)\right ] e^{- \gamma T_i}\nonumber \\
&-\left [A^{22}_{\rm init,i} +i \sqrt{3 \pi/10} \frac{M_*}{n a^3_r\omega}Q_\xi (P_{-2}-P_2) \right] \omega  \sin (\omega T_i)  e^{- \gamma T_i}\,,
\end{align}
with $A^{22}_{\rm init, i} =A^{2,-2 *}_{\rm init, i}, \dot{A}^{22}_{\rm init, i} =\dot{A}^{2,-2 *}_{\rm init, i}$ and similarly
\begin{align}\label{eqa0map}
A^{20}_{\rm init, i+1} & = A^{20}_{\rm init,i} \left [ \cos (\omega T_i) +\frac{\gamma}{\omega} \sin (\omega T_i)\right ] e^{- \gamma T_i}\nonumber \\
&+ \left [-\sqrt{4 \pi/5} \frac{M_*}{n a^3_r}Q_\xi P_0+\dot{A}^{20}_{\rm init,i } \right ] \frac{ \sin (\omega T_i)}{\omega} e^{- \gamma T_i}\,,\nonumber \\
\dot{A}^{20}_{\rm init, i+1} & =\left [-\sqrt{4 \pi/5} \frac{M_*}{n a^3_r}Q_\xi P_0+\dot{A}^{20}_{\rm init,i } \right ]  \left [ \cos (\omega T_i) -\frac{\gamma}{\omega}\sin (\omega T_i)\right ] e^{- \gamma T_i}\nonumber \\
&-A^{20}_{\rm init,i}\omega  \sin (\omega T_i)  e^{- \gamma T_i}\,,
\end{align}
where we have abbreviated the mode index for frequencies and decay rates because they are the same for all modes with $\ell=2$. 

\end{widetext}

\begin{figure}[tb]
\includegraphics[width=8.4cm]{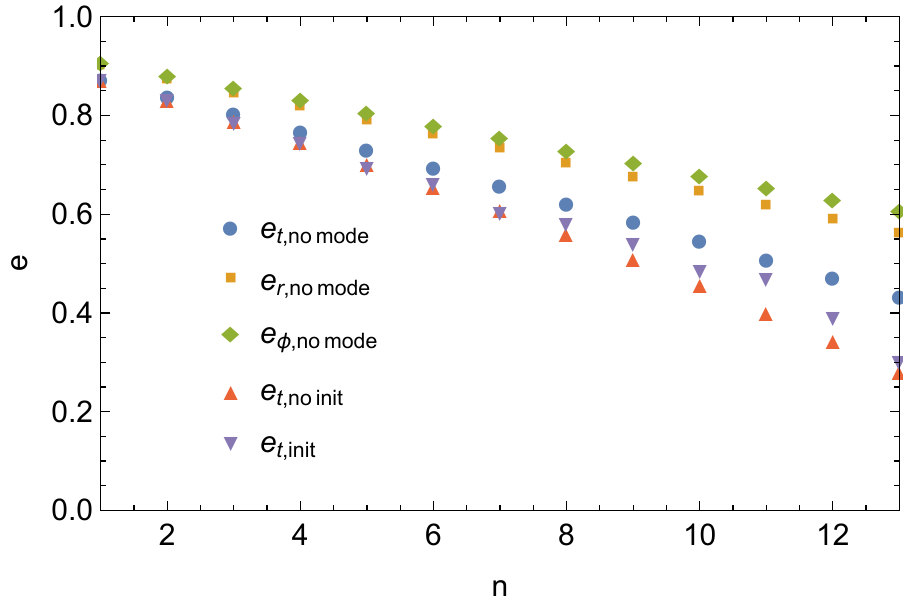}
\caption{Plot of PN eccentricities for a sequence of pericenter passages,
indexed by $n$. The quantities $e_t$, $e_r$, and $e_\phi$ are not gauge-invariant and
their values  are given in modified harmonic coordinates. The initial pericenter
separation is $R_{0,1}=3 R_*$, and the initial $e_r$ is set to $0.9$. The quantities $e_{\rm
t, no \,mode}$, $e_{\rm r, no\, mode}$, and $e_{\rm \phi, no \, mode}$ are
obtained without including the star's oscillations; $e_{\rm t, no \, init}$ is
computed including the star's oscillation but zero $A_{\rm init}$; $e_{\rm t,
init}$ is computed taking into account the evolution of the mode amplitude over
time.}
\label{fig:eplot}
\end{figure}

In reality, even if the initial mode oscillation is known at the beginning 
of a sequence of bursts, the phase error due to the PN approximation and the 
osculating orbit approximation may accumulate with time and eventually lead 
to very inaccurate predictions for $A_{\rm init}$ and $\dot{A}_{\rm init}$. 
As a result, if we denote the time for a template to receive $\mathcal{O}(1)$
phase error as $\tau_1$, an accurate template should satisfy $\tau_1 \gamma_{22}> 1$.

To 1PN order, the periastron frequency of the $i$th cycle $f_i$ is given by \cite{loutrel2017eccentric}
\begin{align}
f_i  & \equiv \frac{1}{2 \pi} \frac{\rm pericenter \,\,\,velocity}{\rm pericenter \,\,\, distance}\, \nonumber \\
& =\frac{1}{2 \pi} \sqrt{\frac{1+e_\phi}{1-e_\phi}} \frac{n K}{1-e_t}\,.
\end{align}

At 3 PN order, $f_i$ can be obtained by combining Eq.~ ($346$) and Eq.~($343$) in \cite{blanchet2014gravitational}:
\begin{widetext}
\begin{align}
f_i & =\frac{n}{2\pi}\frac{d \phi/d u}{d \ell/du} =\frac{n K}{2\pi} \left . \frac{dv}{du} \right |_{u=0} \left .\frac{1+2 f_\phi \cos 2v +3 g_\phi \cos 3 v + 4 i_\phi \cos 4 v +5 h_\phi \cos 5 v}{1-e_t \cos u-g_t+ dv/du(g_t+f_t \cos v+2 i_t \cos 2v+3 h_t \cos 3v)} \right |_{u,v=0} \nonumber \\
& = \frac{n K}{2 \pi} \sqrt{\frac{1+e_\phi}{1-e_\phi}} \frac{1+2 f_\phi  +3 g_\phi  + 4 i_\phi +5 h_\phi }{1-e_t -g_t+ \sqrt{\frac{1+e_\phi}{1-e_\phi}}(g_t+f_t +2 i_t +3 h_t )}\,,
\end{align}
where the definition and more discussions on the orbital elements  $f_\phi, g_\phi, i_\phi, h_\phi, f_t, g_t, i_t, h_t, e_t, e_r, e_\phi$ can be found in \cite{blanchet2014gravitational,memmesheimer2004third}.

Under the impulsive approximation, the Newtonian formula for the energy and angular momentum deposited into  the stars, i.e., Eq.~\eqref{eqeinitd} and Eq.~\eqref{eqjinitd}, can be further improved by incorporating the 1PN trajectory description in Eq.~\eqref{eqtra1pn}. To do that, we only need to replace $a_0$ in Eq.~\eqref{eqjinitd}, \eqref{eqeinitd} by $a_r$, and the definition of $P_m(y)$ to be
\begin{align}
P_{m}(y)  = \int^{ \pi}_0  \, \frac{dx}{(1-e_r \cos x)^2} 
 \times \cos {\left \{ y \frac{x -e_t \sin x}{n}+ 2 m K \arctan \left [ \left ( \frac{1+e_\phi}{1-e_\phi}\right )^{1/2} \tan \frac{u}{2}\right ]\right \} }\,,
\end{align}
and $\Delta E_{\rm mode}$ and $\Delta J_{\rm mode}$ are explicitly given by

\begin{align}\label{eqjinitpd}
 \Delta J_{\rm mode,z} &= |Q^{22}_\xi|^2 \frac{3 \pi}{5 \omega_{22}} \frac{M^2_*}{n^2 a^6_r}[P^2_{-2}(\omega_{22})-P^2_{2}(\omega_{22})]
+ \frac{2 \sqrt{6 \pi/5} M_*}{a^{3}_r n} [P_{-2}(\omega_{22})+P_{2}(\omega_{22})] {\rm Im}[A^{(22)}_{\rm init} Q^{22}_\xi] \nonumber \\
&+ \frac{2 \sqrt{6 \pi/5} M_*}{\omega_{22} a^{3}_r n} [P_{-2}(\omega_{22})-P_{2}(\omega_{22})] {\rm Re}[\dot{A}^{(22)}_{\rm init} Q^{22}_\xi] \,,
   \end{align}
and
 \begin{align}\label{eqeinitpd}
  \Delta E_{\rm mode} & = |Q^{22}_\xi|^2 \frac{3 \pi}{5} \frac{M^2_*}{n^2 a^6_r}[P^2_{-2}(\omega_{22})+P^2_{2}(\omega_{22})]
 +|Q^{20}_\xi|^2 \frac{2 \pi}{5 } \frac{M^2_*}{n^2 a^6_r} P^2_{0}(\omega_{20})\, \nonumber \\
 &+ \frac{ \omega_{22} \sqrt{6 \pi/5} M_*}{ a^{3}_r n} [P_{-2}(\omega_{22})-P_{2}(\omega_{22})] {\rm Im}[A^{(22)}_{\rm init} Q^{22}_\xi] 
 +\frac{ \sqrt{6 \pi/5} M_*}{ a^{3}_r n} [P_{-2}(\omega_{22})+P_{2}(\omega_{22})] {\rm Re}[\dot{A}^{(22)}_{\rm init} Q^{22}_\xi] \nonumber \\
 &- \frac{2\sqrt{\pi/5} M_*}{a^{3}_r n}\dot{A}^{(20)}_{\rm init} Q^{20}_\xi P_0(\omega_{20})\,.
      \end{align} 
      
\end{widetext}

\begin{figure}[tb]
\includegraphics[width=8.4cm]{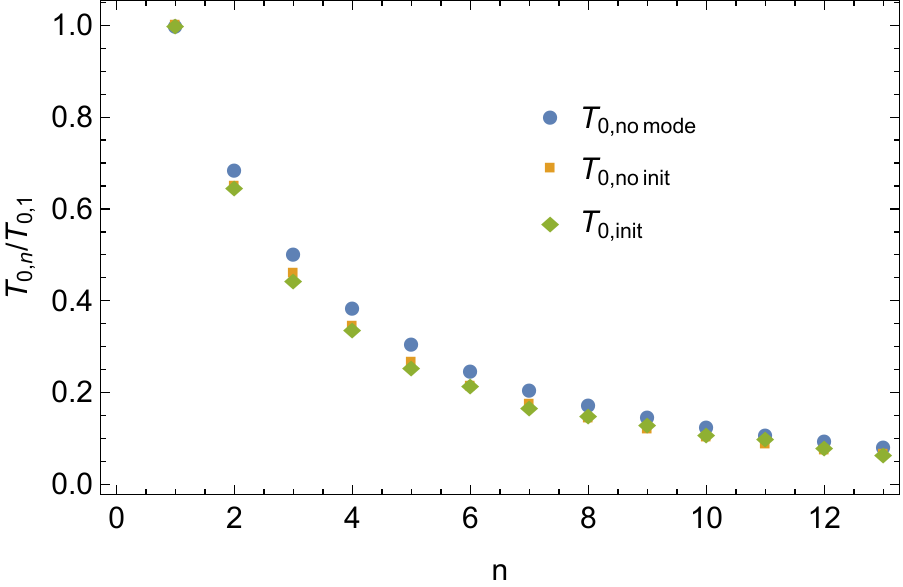}
\includegraphics[width=8.4cm]{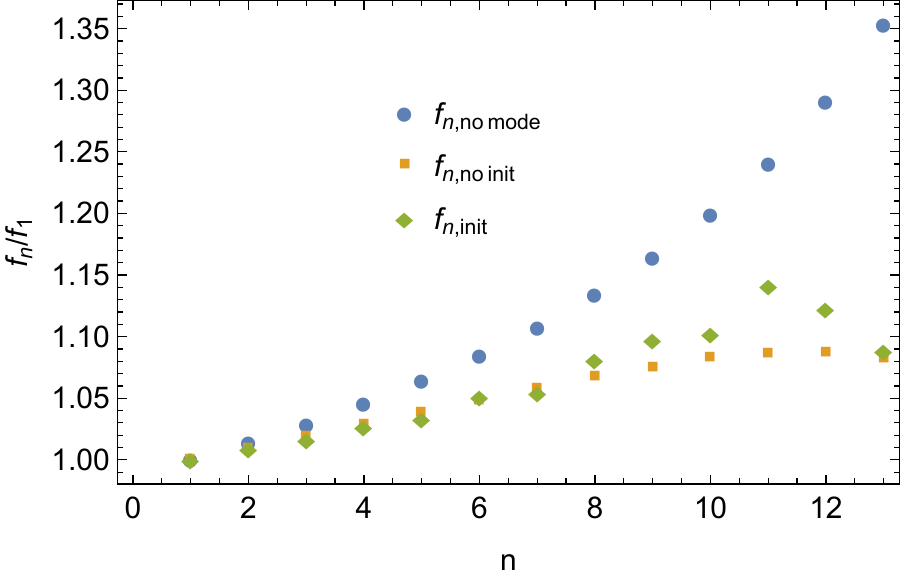}
\includegraphics[width=8.4cm]{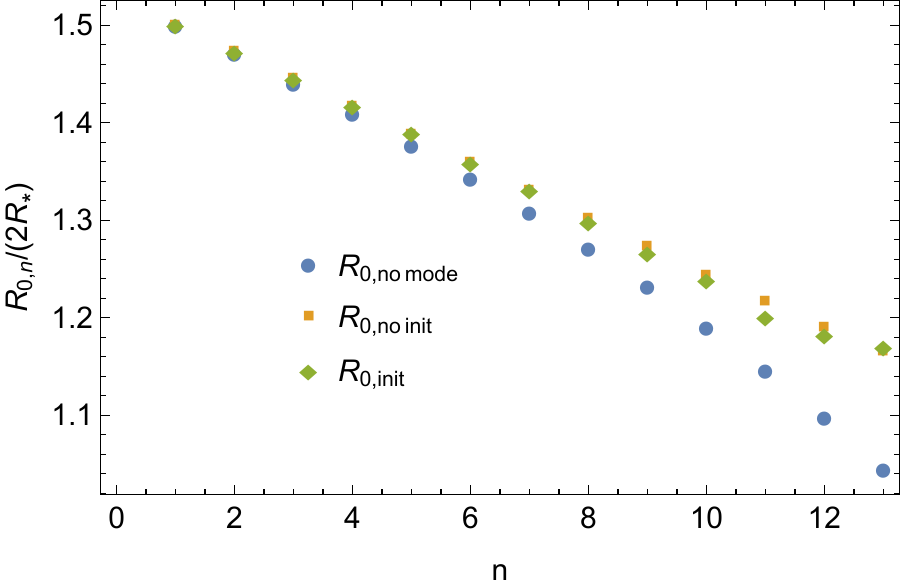}
\caption{The same set-up as Fig.~\ref{fig:eplot}. The top plot represents the evolution of the orbital period; the middle panel represents the evolution of the pericenter frequency; the bottom panel represents the evolution of the pericenter distance.}
\label{fig:TfRplot}
\end{figure}
Note that in order to capture  higher PN tidal effects, the analysis on stellar oscillation in Appendix A should be extended to higher PN orders as well.
In order to illustrate the trajectory model described by
Eqs.~\eqref{eqjinitd},~\eqref{eqeinitd},~\eqref{eqa2map},~\eqref{eqa0map},~\eqref{eqEJmap},~\eqref{eqDEJmap},
we start an orbit with initial pericenter distance $R_0 = 3 R_*$ and initial
eccentricity $e_r =0.9$ (in the modified harmonic coordinate; this quantity is not
gauge-invariant). In Fig.~\ref{fig:eplot}, we compare the evolution of
eccentricities in three different scenarios: (i) assuming no tidal excitations of the stars,
(ii) allowing stellar oscillations, while neglecting the non-zero values of 
$A_{\rm init}$ and $\dot{A}_{\rm init}$ from previous encounters, (iii) allowing star
oscillations and evolving $A_{\rm init}$ and $\dot{A}_{\rm init}$.
In Fig.~\eqref{fig:TfRplot}, we also present the evolution of  
the orbital period, pericenter frequency, and pericenter distance for the above
three scenarios. We can see that for such a close encounter, the oscillation of the
stars significantly alters the trajectory, and it is important to include the
evolution of the modes into the trajectory model. 
In fact, for this case during some of the later close encounters, the frequency of the
orbit is larger when this evolution is tracked, compared to when it is not,
as energy and angular momentum are taken out
of the NS oscillations and put back in the orbit.  
We also notice that, for the last
several orbits, the eccentricity falls below $0.5$. Strictly speaking, orbits
in such a regime are no longer highly eccentric, since the impulsive
approximation is no longer accurate. To deal with orbits with mid-range
eccentricities, the mode evolution has to be computed using a continuous forcing
in time. Studying this regime is beyond the scope of this paper.

\section{Conclusion}
In this work, we have shown that tidal excitation of stellar modes can dramatically
influence the dynamics of binary neutron stars in highly eccentric orbits.
The amount of extra energy and angular momentum
change in the orbit due to mode excitation can be tens of percent
of that due to GWs alone, if the pericenter distance is smaller
than $\sim 4 R_*$. The exact amounts will depend on the EOS, 
and here we focused on a $\Gamma=2$ polytrope
to simplify the comparison between the analytic model and
numerical results.
The prediction from the Newtonian approximation
agrees with the full numerical results to within a factor of
a two (with the error likely dominated by systematic effects in measuring
orbital properties from the code results).
This also resolves the large discrepancy between a similar analytic calculation and
numerical simulations reported in \cite{chirenti2016gravitational}.

The discussion presented in this work also applies for highly eccentric NS-BH
binaries, regardless of the mass ratio \cite{yang2018can}. While the rates of
detecting these highly eccentric binaries are uncertain, their gravitational
waveforms display distinctive f-mode oscillations, which are not present
in any appreciable amount in quasi-circular systems.  Observing these oscillations
would provide unprecedented information about the structure and EOS of cold NSs.
One possible way to enhance the SNR of these oscillation features is to
stack the post-encounter waveform of a series of pericenter passages, which
requires accurate predictions for the timing of these encounters. Another
possibility is to rely on the observation of third-generation ground-based
detectors. As the f-mode frequency is generally above 2 kHz (the  f-mode
frequency in Fig.~\ref{fig:bhns_nsns_gw} is approximately 2.6 kHz), the
high frequency detector proposed in \cite{miao2017towards} is ideal for 
observing such signals. Assuming a detection strategy that coherently adds the signals from different pericenter encounters, the estimated SNR for f-mode oscillations is
\begin{align}
{\rm SNR} \sim & \,30 \left ( \frac{\mathcal{E}_{mode}}{\mathcal{E}_{\rm GW}}\right )^{1/2} \left( \frac{50 {\rm \ Mpc}}{d}\right ) \left ( \frac{5 \times 10^{-25} {\rm \ Hz}^{-1/2}}{\sqrt{S_n}}\right ) \nonumber \\ & \times \left ( \frac{2000 {\rm \ Hz}}{f}\right )\,,
\end{align}
where $d$ is the distance of the binary from Earth, $S_n$ is the one-sided power spectral density of the detector, $\mathcal{E}_{\rm GW}$ is the total energy radiated by GWs and $\mathcal{E}_{mode}$ is the energy radiated by f-mode oscillations.

\acknowledgments %
The authors thank Nathan K. Johnson-McDaniel for interesting discussions.  F.P., V.P.,
H.Y. ~acknowledge support from NSF grant PHY-1607449, the Simons
Foundation, NSERC and the Canadian Institute For Advanced Research (CIFAR). 
V.P. also acknowledges support from NASA grant
NNX16AR67G (Fermi). This research was supported in part by the Perimeter
Institute for Theoretical Physics. Research at Perimeter Institute is
supported by the Government of Canada through the Department of
Innovation, Science and Economic Development Canada and by the
Province of Ontario through the Ministry of Research, Innovation and
Science.  Computational resources were provided by XSEDE under grant
TG-PHY100053 and the Perseus cluster at Princeton University.

\appendix
\section{Perturbations of a polytropic star}\label{appendixA}

The perturbation of a polytropic star is described by the Lagrangian displacement field ${\bf \xi}$, which obeys (Eq.~(2.212) of \cite{poisson2014gravity}):
\begin{align}\label{eqeomp}
\partial^2_t \xi =\left ( \frac{\delta \rho}{\rho^2}\right ) \nabla \rho -\frac{\nabla \delta p}{\rho} -\nabla \delta \Phi\,,
\end{align}
where
\begin{align}
\delta p & = - \Gamma_p p \nabla \cdot \xi -\xi \cdot \nabla p \,\nonumber\\
\delta \rho & = -\nabla \cdot (\rho \xi)\,.
\end{align}
Here $\Gamma_p$ is the adiabatic index of the perturbation, which may or may not be the same as $\Gamma$ for the equilibrium configuration. However, for the polytropic (hence barotropic) stars studied here, we have $\Gamma=\Gamma_p$.  In addition, the total Newtonian potential is $\delta \Phi =\delta U + U_{\rm tide}$, where $\delta U$ obeys the Poisson equation:
\begin{equation}\label{eqpoi}
\nabla^2 \delta U = -4 \pi G \delta \rho\,.
\end{equation}

The equation of motion for the Lagrangian displacement field can be written schematically as
\begin{align}\label{eqeomm}
\partial^2_t \xi_j +{L_j}^k \xi_k =\nabla_j U_{\rm tide}\,.
\end{align}
In order to determine the body's response to an applied tidal field, it is
useful to first compute the normal modes of the system, corresponding to free
oscillations, i.e. solutions of Eq.~\eqref{eqeomm} with the right-hand side being
zero. With $\xi^{(n)}$ given in Eq.~\eqref{eqeigenwf}, we write $\xi_j$ as
$\xi^{(n)}_j e^{i \sigma t}$, and define
\begin{align}\label{eqp}
\delta p & = \rho(r)\, y(r) Y_{lm}(\theta, \phi) e^{i \sigma t}\,,\nonumber\\
\delta \rho & = \varrho(r) Y_{lm}(\theta, \phi) e^{i \sigma t} \,,\nonumber \\
\delta U & = u(r) Y_{lm}(\theta, \phi) e^{i \sigma t}\,.
\end{align}
When combining Eqs.~\eqref{eqeomp} and \eqref{eqpoi}, we notice that not all
perturbation variables are independent. In particular, we can express $\xi_S$
and $\varrho$ in terms of the other variables as 
\begin{align}
\xi_S = \frac{y+u}{r \sigma^2}\,,\quad \varrho = \frac{\rho^2 y}{\Gamma_p p}-A \xi_R\,.
\end{align}
We obtain the following ordinary differential equations:
\begin{align}\label{eqode}
(r^2 \xi_R)' & = -\frac{r^2 p' \xi_R}{\Gamma_p p}+\left [ \frac{l(l+1)}{\sigma^2} - \rho \frac{r^2}{\Gamma_p p}\right ] y+ \frac{l(l+1) u}{\sigma^2}\,,\nonumber\\
y' & = (\sigma^2+A g) \xi_R-A y-u'\,,\nonumber\\
u'' & = -\frac{2}{r } u'+\frac{l(l+1) u}{r^2}+4 \pi G \rho \left ( \frac{\rho y}{\Gamma_p p} - A \,\xi_R \right )\,
\end{align}
with
\begin{align}
g \equiv - \frac{p'}{p},\quad A \equiv \frac{\rho'}{\rho} -\frac{p'}{\Gamma_p p}\,.
\end{align}
As explained above, for the polytropic stars studied here, $A$ is zero. The
differential equations are subject to the regularity condition at the center of star
and boundary conditions at the stellar surface which require force balance and
zero pressure:
\begin{align}
& (\rho \,y+ p' \xi_R) |_{R_*} =0,\nonumber \\
 & \left. \left ( u'+(l+1)\frac{u}{r} \right ) \right|_{R_*} =-4 \pi G \rho(R_*) \xi_R(R_*)\,.
\end{align}
The eigenfrequency $\sigma$ and eigenfunctions $\xi_R$ and $\xi_S$ can be obtained by solving 
Eqs.~\eqref{eqode} with the above boundary conditions.

\bibliography{master}
\end{document}